\documentclass[a4paper,fleqn]{cas-dc}

\usepackage[numbers]{natbib}
\usepackage{amsmath,amssymb}
\usepackage{algorithm}
\usepackage{algorithmic}
\usepackage{soul}
\usepackage{amsmath}
\usepackage{nomencl}
\usepackage[switch]{lineno} 
\makenomenclature

\def\tsc#1{\csdef{#1}{\textsc{\lowercase{#1}}\xspace}}
\tsc{WGM}
\tsc{QE}

\begin{document}
\let\WriteBookmarks\relax
\def\floatpagepagefraction{1}
\def\textpagefraction{.001}

\title [mode = title]{Life cycle economic viability analysis of battery storage in electricity market}  



%
\author[1]{Yinguo Yang}[]
\author[2]{Yiling Ye}[]






\author[2]{Zhuoxiao Cheng}[]




\author[2]{Guangchun Ruan}[]
\author[1]{Qiuyu Lu}
\author[2]{Xuan Wang}[]
\author[2]{Haiwang Zhong}[]






\affiliation[1]{organization={Guangdong Power Grid Corp},
            city={Guangdong},
            country={China}}
\affiliation[2]{organization={Department of Electrical Engineering, Tsinghua University},
            city={Beijing},
            postcode={100084}, 
            country={China}}
       

\begin{abstract}
Battery storage is essential to enhance the flexibility and reliability of electric power systems by providing auxiliary services and load shifting. Storage owners typically gains incentives from quick responses to auxiliary service prices, but frequent charging and discharging also reduce its lifetime. Therefore, this paper embeds the battery degradation cost into the operation simulation to avoid overestimated profits caused by an aggressive bidding strategy. Based on an operation simulation model, this paper conducts the economic viability analysis of whole life cycle using the internal rate of return(IRR). A clustering method and a typical day method are developed to reduce the huge computational burdens in the life-cycle simulation of battery storage. Our models and algorithms are validated by the case study of two mainstream technology routes currently: lithium nickel cobalt manganese oxide (NCM) batteries and lithium iron phosphate (LFP) batteries. Then a sensitivity analysis is presented to identify the critical factors that boost battery storage in the future. We evaluate the IRR results of different types of battery storage to provide guidance for investment portfolio.
\end{abstract}




\maketitle

\printnomenclature

\nomenclature{$k$}{Used as a subscript to indicate half cycle of battery charging and discharging.}
\nomenclature{$C$}{The set of battery charging and discharging half cycle.}
\nomenclature{$e$}{Used as a superscript to denote the energy market.}
\nomenclature{$res$}{Used as a superscript to denote the spinning reserve ancillary service market.}
\nomenclature{$reg$}{Used as a superscript to denote the frequency regulation ancillary service market.}
\nomenclature{$ch$}{Used as a superscript to denote battery charging.}
\nomenclature{$dch$}{Used as a superscript to denote battery discharging.}
\nomenclature{$cap$}{Used as a superscript to denote frequency regulation capacity.}
\nomenclature{$\textit{perf}$}{Used as a superscript to denote frequency regulation performance.}
\nomenclature{$\textit{op}$}{Used as a superscript to denote operation and maintenance cost of battery storage.}
\nomenclature{$\textit{loss}$}{Used as a superscript to denote battery degradation.}
\nomenclature{$\textit{fix}$}{Used as a superscript to denote the fixed part of income or cost.}
\nomenclature{$\textit{var}$}{Used as a superscript to denote the variable part of income or cost.}
\nomenclature{$\textit{LFP}$}{Used as a subscript to indicate lithium iron phosphate (LFP) batteries.}
\nomenclature{$\textit{NCM}$}{Used as a subscript to indicate lithium nickel cobalt manganese oxide (NCM) batteries.}
\nomenclature{$\textit{mkt}$}{Used as a subscript to indicate benefits of battery storage from participating in the electricity market.}
\nomenclature{$\textit{rcy}$}{Used as a subscript to indicate battery recycling revenue.}
\nomenclature{$\textit{bat}$}{Used as a subscript to indicate battery purchasing cost.}
\nomenclature{$\textit{equ}$}{Used as a subscript to indicate equipment purchasing cost.}
\nomenclature{$\textit{sta}$}{Used as a subscript to indicate power station design and construction cost.}
\nomenclature{$\Delta t$}{Time unit.}
\nomenclature{$\textit{$P_r$}$}{The rated power of battery storage (MW).}
\nomenclature{$\textit{$E_r$}$}{The rated energy of battery storage (MWh).}
\nomenclature{$E_{\textit{max/min}}$}{The upper/lower limit of the energy held in battery storage (MWh).}
\nomenclature{$\textit{e}_\textit{t}$}{The state of charge at hour t, i.e. the enegy held in the battery storage at hour t.}
\nomenclature{\textit{$\Delta e_t$}}{The amount of energy change in hour $t$.}
\nomenclature{$k_{\textit{fix}}$}{The empirical unit values of the fixed part of the operation and maintenance cost.}
\nomenclature{$k_{\textit{var}}$}{The empirical unit values of the variable part of the operation and maintenance cost.}
\nomenclature{$\textit{Cap}^{\textit{loss}}$}{The battery capacity loss.}
\nomenclature{$\textit{T}$}{Temparature of the battery.}
\nomenclature{$\textit{d}$}{The specific value of the depth of discharge.}
\nomenclature{$N_d$}{The maximum number of cycles before the battery reaches the end of its life when the depth of discharge is d.}
\nomenclature{$e^{\textit{reg}}$}{The average hourly charge or discharge energy per MW frequency regulation capacity, MWh.}
\nomenclature{$\textit{Prob}^{\textit{res}}$}{The probability of spinning reserve's deployment.}
\nomenclature{$r_\textit{self}$}{The battery self-discharge rate.}
\nomenclature{$\textit{LCOS}$}{The levelized cost of storage.}
\nomenclature{$\textit{FEC}$}{Full equivalent cycles, which is the average of battery charge and discharge cycles for the given period of time.}
\nomenclature{$\textit{DoD}$}{Depth of discharge, which is the percentage of the battery that has been discharged relative to the overall capacity of the battery.}
\nomenclature{$\textit{SoC}$}{State of Charge, which is a measurement of the amount of energy available in a battery at a specific point in time expressed as a percentage.}
\nomenclature{$_t$}{Used as a subscript to represent hours or year.}
\nomenclature{$_T$}{Used as a subscript to represent the incomes or costs by the end of year $T$.}
\nomenclature{$_i$}{Used as a subscript to represent dates.}
\nomenclature{$M$}{Set of typical days.}
\nomenclature{$\textit{IRR}$}{Tnternal rate of return, which is a metric used in financial analysis to estimate the profitability of potential investments.}
\nomenclature{$\textit{LMP}$}{Locational Marginal Pricing, which is a way for wholesale electric energy prices to reflect the value of electric energy at different locations.}
\nomenclature{$\textit{SOH}$}{State of health, which describes the overall health and remaining capacity of a battery.}
\nomenclature{$\textit{C}_\textit{ch}$}{Battery charging rate.}
\nomenclature{$\textit{C}_\textit{dch}$}{Battery discharging rate.}
\nomenclature{\textit{$mSOC$}}{Mean State of Charge.}

\section{Introduction}\label{section1}
Energy balance of modern power systems becomes increasingly challenging when a high penetration of renewable energy is required to fulfill the ambitious goal of carbon neutrality~\cite{shan2021role}. Cost effective energy storage such as battery storage is urgently needed to provide flexibility resources to accommodate the intermittent renewable energy~\cite{ruan2021estimating}. 

Battery storage is highly valuable in the ancillary service market and the energy market. In the ancillary market, battery storage is favored for its rapid response, which is widely applied in ancillary services including frequency regulation and spinning reserve. For frequency regulation, since significant frequency fluctuations may threaten the stability of the power grid, storage owners are urged to adjust the electricity output according to the frequency regulation signal to compensate for the imbalance between supply and demand in order to keep the frequency within a safe range. RegA is the traditional regulation signal commonly used in frequency regulation market, which is obtained from the regional control error after passing through a low-pass filter. However, RegA(traditional) with a sampling time of 15 minutes ignores the rapid response potentials of battery storage when applied for frequency regulation. Therefore, a dynamic regulation signal RegD sent every 2 seconds will be a more reasonable choice which rewards the ability of quick response\cite{PJMReg}. A high-pass filter instead of the low-pass one is used to obtain RegD(dynamic). Since the output power of the battery storage fluctuates according to the regulation signal, the concept of regulation mileage is introduced to measure the movement of battery storage output power. The performance of battery storage is scored according to its accuracy of response to the regulation signal. Reference \cite{rangarajan2023assessing} concluded that grid-scale batteries can significantly lower overall frequency regulation costs through analysis based on Australian electricity markets.

For spinning reserve, storage owners are required to respond to the instructions of the system operator in a very short time in case of sudden power supply shortages. Therefore, battery storage which can adjust the discharge power in seconds is truly valued. 

Besides, in the energy market, since the supply and the demand of electricity must be strictly cleared all the time, load variations across different time periods cause time-varying electricity price. Therefore, battery storage can also benefit from load shifting in the energy market. Reference \cite{tang2023life} suggested that battery storage could be a profitable choice in building-scale investment in the future due to downward trends in capacity costs and more energy flexibility demand of power system. Reference \cite{cao2020deep} adopted a model-free deep reinforcement learning (DRL) method to optimize the battery energy arbitrage considering an accurate battery degradation model, which recognizes the profitability of battery storage in the energy market. More definitions and acronyms about terms in electricity market can be found in the glossary of reference \cite{caiso_glossary}.

However, the high cost at present and unclear market returns are hindering the large-scale application of battery storage systems. It is thus essential to evaluate the life cycle economic viability of battery storage in an accurate and comprehensive way~\cite{walawalkar2007economics}.

In the literature, there are two methodologies for analyzing life cycle economic viability: the levelized cost of storage (LCOS) analysis and the optimal bidding strategy considering the lifetime of battery storage.

Based on the concept of levelized cost of electricity, LCOS can be defined as the total lifetime cost of the investment in an electricity storage technology divided by its cumulative delivered electricity\cite{pawel2014cost,julch2016comparison}, shown as:

\begin{subequations}
    \label{eq:LCOS}
\begin{align}
    \textit{LCOS}&=\frac{\textit{CAPEX}+\sum_{t=1}^{t=n}\frac{A_t}{(1+i)^t}}{\sum_{t=1}^{t=n}\frac{W_{out,t}}{(1+i)^t}}\\
    A_t&=\textit{OPEX}_t+\textit{CAPEX}_{re,t}+c_{el}\cdot W_{in,t}-R_t
\end{align}
\end{subequations}
where CAPEX refers to the capital expenditure while the annual cost of the storage system $A_t$ at each point of time t is added up over the lifetime n of the storage. $W_{out,t}$ indicates the annual output energy. A discount rate \textit{i} is considered to convert $A_t$ and $W_{out,t}$ to the same point of time. Reference \cite{julch2015holistic} and \cite{abdon2017techno} also detailed how to calculate LCOS, in which environmental factors such as carbon emissions were also considered. Based on similar concepts, LCOS of photovoltaic power plants equipped with battery storage was investigated in \cite{pawel2014cost} and \cite{lai2017levelized}. Reference~\cite{zakeri2015electrical} and \cite{schmidt2019projecting} considered the impact of different application scenarios, and it was found in \cite{schmidt2019projecting}  that lithium-ion batteries would become the most economical option among all technical routes. Reference~\cite{cristea2022levelized} highlighted the LCOS difference between countries caused by different market policies and conducted a market research considering the effect of subsidies offered within the Romanian programs instead of directly using literature data of LCOS, in which Lead-carbon (LCB) and Lithium-ion (Li-ion) were found to be the best and second best profitable battery technology used by household and industrial consumers as storage systems in Romania. In power-type energy storage applications, \cite{shi2017using} calculated not only battery storage cost per kilowatt-hour, but also that per mileage corresponding to mileage compensation in the electricity market. In the LCOS method, the capacity decay of battery storage is simplified by taking the average value, which results in relatively low accuracy. Reference~\cite{steckel2021applying} concerned about the second-life batteries and  presented potential improvements to the LCOS model for second-life batteries. 

Existing literature has rich discussion on the optimal bidding issue under various market conditions and different assumptions~\cite{qiu2022strategic,de2016optimal,ruan2021constructing}. However, most of these works rarely considered the decay of battery storage capacity caused by frequent charge and discharge cycling, resulting in an aggressive bidding strategy~\cite{akhavan2013optimal, kazemi2017operation}. Given that this extra cost of battery storage can be significant under frequent regulation service, degradation cost was introduced in \cite{xu2017factoring} to ensure an optimal solution. Reference \cite{reniers2021unlocking} demonstrated that applying physics-based battery degradation model can increase revenue and decrease degradation of battery storage in comparison with the linear model, which proves the importance of considering battery degradation of battery storage. Reference~\cite{maheshwari2020optimizing}  demonstrated that degradation in lithium-ion batteries is non-linearly dependent on the operating conditions and proposed a decomposition technique to give near-optimal results for longer time horizons. Weights of $\omega$ and (1-$\omega$) were assigned to the profit and battery degradation cost respectively in the objective function, allowing users to flexibly adjust the importance of profit and battery degradation in their battery storage bidding strategies. Reference \cite{pandvzic2020optimal} considered battery storage as a price maker in reserve market and presented a bilevel model for optimal battery storage participation in day-ahead energy market and reserve market. The proposed bilevel model was converted into a mixed-integer linear program(MILP) by using the Karush–Kuhn–Tucker optimality conditions. Reference \cite{knevzevic2020optimal} analyzed the optimal bidding of virtual prosumer on day-ahead market, which is the integration a group of prosumers.

According to the three levels of lithium-ion batteries including the material/electrode level, the cell level, and the module/pack level, Lithium-ion battery degradation model can be divided into three categories. Reference \cite{li2019development}  developed a single particle model at the electrode level to predict the lifetime of grid-connected lithium-ion battery energy storage system with great accuracy. Reference \cite{mishra2021model} developed a detailed and efficient physics-based cell-level degradation model which coupled temperature gradients within the battery. Considering that the battery degradation model at the material/electrode level is accurate but too complex while the additional degradation at the module/pack level will be generally controlled within an acceptable range, the battery degradation model at the cell level was extensively adopted in the solution of an optimization problem, which best balances accuracy and computational complexity. 

According to the different fitting  adopted, the Ah-based semi-empirical degradation model based on Arrhenius law and the cycle-based model based on Miner's Rule are mainly applied to estimate the cycle degradation of battery energy storage systems. 

Lifetime prediction was conducted in \cite{schmalstieg2014holistic} and \cite{ecker2012development} based on Ah-based model. Reference \cite{zhang2021improved} applied the Ah-based model and transformed the issue into mixed-integer linear programming (MILP) to simplify the calculation. With the application of the Ah-based model, reference \cite{wang2014degradation} analyzed the cycling-induced capacity decay in NCM + LMO/graphite Li-ion cells, while a LiFePO4 battery was investigated in \cite{wang2011cycle}. Reference \cite{petit2016development} used experimental data to consider the influence of V2G on LFP/C and NCA/C batteries.

The cycle-based model firstly uses full equivalent cycles (FEC) to describe battery degradation, and then obtains FEC's function $\textit{f(FECs)}$, while the influence of depth of discharge (DoD), i.e. the State of Charge (SoC) difference in a cycle, and other contributing factors are taken into account as the coefficients of $\textit{f(FECs)}$. DoD was generally considered as the main factor in many articles\cite{he2015optimal,naumann2020analysis,laresgoiti2015modeling,xu2017factoring}. Reference \cite{shi2017using} proposed that the degradation ratio of battery storage was linearly related to DoD and simplified the model, while the influence of other factors was usually ignored. The rainflow counting algorithm\cite{xu2017factoring} is extensively used to calculate the number of cycles in the SoC Profile and the DoD of each cycle, which is following the same idea with FEC(i.e. the number of cycles when DoD is 100\%). Reference \cite{kwon2022reinforcement} developed a cycle-based battery degradation model which adopted the rainflow algorithm using the powerful reinforcement learning algorithms. A gradient descent algorithm to quickly solve battery degradation optimization problems was conducted in \cite{shi2018convex}. 

No matter which fitting formula is adopted, the semi-empirical models at the cell level in the aforementioned research are mostly based on the experimental data of a specific type of battery, which leads to poor universality. In response to this problem, reference \cite{olmos2021modelling} synthesized all the existing degradation experimental data and conducted two degradation models suitable for lithium nickel cobalt manganese oxide (NCM) batteries and lithium iron phosphate (LFP) batteries respectively.

In brief, LCOS is the method commonly used for the life cycle economic viability analysis of battery storage, yet its accuracy is limited since it only roughly approximates the impact of battery degradation and electricity price fluctuations. The optimal bidding strategy model considering the lifetime of battery storage takes battery degradation into account, yet it is applied mostly to guide the day-ahead bidding of battery storage and has not been applied to the life cycle economic viability analysis of battery storage.

To solve the issues addressed above, this paper proposes a life cycle economic viability analysis model based on the operation simulation of battery storage. The main contributions of this paper are listed as follows:

\begin{itemize}
\item[1)] This paper simulates the battery storage daily operation considering its participation in frequency regulation, spinning reserve and load shifting. An optimization problem is solved with the income of battery storage maximized to obtain the operation simulation result. The battery degradation cost is taken into account to avoid aggressive bidding strategy.

\item[2)] In the detailed consideration of battery degradation, this paper adopts a model with better universality to LFP and NCM technologies respectively while conducting certain simplifications to balance accuracy and calculation efficiency. Key factors other than DoD are also included in the analysis of degradation, such as temperature, current rate, and average state of charge. 

\item[3)] This paper proposes a life cycle economic viability analysis model for battery storage based on operation simulation of each day in the whole battery life cycle. Through operation simulation, internal rate of return(IRR) of the battery storage project is obtained by calculating the cash flow of each year based on the cost and revenue of battery storage, which is taken as an accurate measurement. Two different methods are applied in the detailed calculation of the IRR, including a clustering method and a method which selects typical days in each year with full-scale consideration to the influence of month, week, festival, etc.
\end{itemize}

The rest of this paper is organized as follow. Section II set up the basic problems in electricity market. Section III formulates the operation simulation model of battery storage considering capacity decay. The life cycle economic viability analysis model of battery storage is proposed in section IV with IRR as a comprehensive indicator. Section V provides the case study results and illustrates the advantage of our model in comparison with LCOS method. Section VI draws the conclusion.

\section{Problem Formulation}\label{section2}

This paper focuses on the life cycle economic viability analysis of battery storage represented by lithium-ion batteries. Without loss of generality, this paper assumes that battery storage mainly provides auxiliary services including frequency regulation and spinning reserve in auxiliary service market, and load shifting in energy market. The US and European electricity markets are mainly considered since they are relatively mature. Considering its relatively small capacity, battery storage is assumed to only act as a price-taker based on data of the day-ahead electricity market.

\begin{figure}[htbp]
    \centering
    \includegraphics[width=0.95\linewidth]{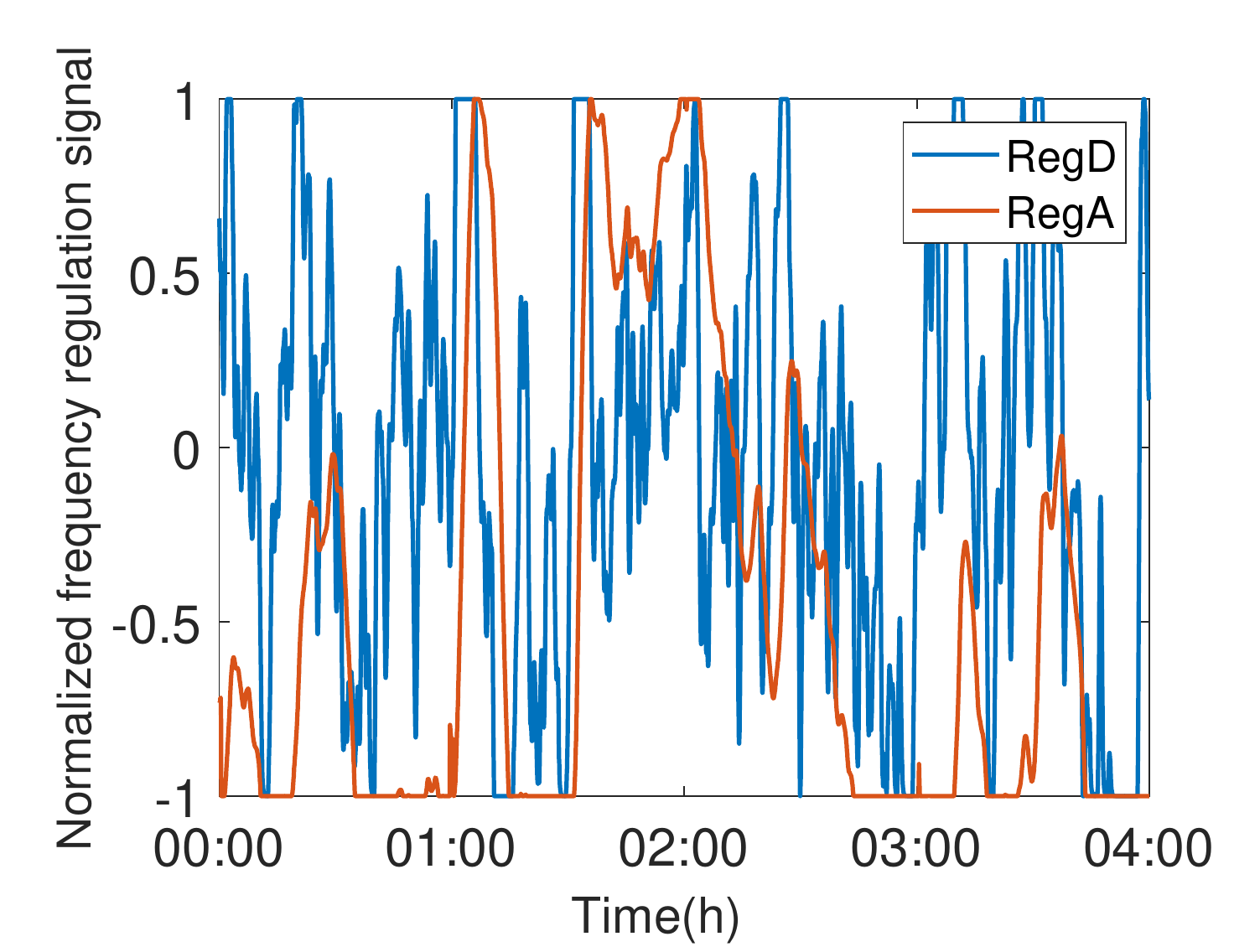}
    \caption{RegA and RegD of PJM electricity market from 0:00 to 4:00 in January 10th, 2020}
    \label{fig:figure_RegA_RegD}
\end{figure}

Battery storage is a regulation resource of higher quality when applied for frequency regulation, which should be incentivized. Therefore, PJM(Pennsylvania-New Jersey-Maryland Interconnection) conducts performance-based regulation, in which RegD(dynamic) with higher fluctuation frequency is introduced on top of RegA(traditional), as shown in figure \ref{fig:figure_RegA_RegD}, which is plotted using the raw data of RegA and RegD of PJM electricity market from 0:00 to 4:00 in January 10th, 2020. 

The payment mechanism of performance-based regulation is shown in formula \eqref{eq:Income_reg1}-\eqref{eq:Income_reg}, which calculates capacity payment $\textit{Income}^\textit{reg,cap}$ and performance payment $\textit{Income}^\textit{reg,perf}$ respectively\cite{he2015optimal}\cite{PJMManual11}.
\nomenclature{\textit{$\textit{Income}^\cdot$}}{The payment for different services that battery storage provides in electricity market(\$).}
\nomenclature{\textit{$\textit{Cost}^\cdot$}}{The cost for different services that battery storage provides in electricity market(\$).}
\nomenclature{$\textit{Price}_t^{(\cdot)}$}{The market price of auxiliary service market and energy market(\$/MWh or \$/MW).}
\nomenclature{$\textit{Cap}_t^{(\cdot)}$}{The capacity of battery storage deployed in auxiliary service market and energy market(MW).}
\nomenclature{$\textit{R}^\textit{mileage}$}{Frequency regulation ratio, which is the mileage of RegD (dynamic) divided by the mileage of RegA (traditional).}
\nomenclature{$\textit{Score}^{\textit{perf}}$}{Score of frequency regulation performance assessment.}
\begin{subequations}
\begin{align}
    \textit{Income}^{\textit{reg,cap}}  =\sum_{t=1}^{24}&\textit{Price}^{\textit{reg,cap}}_t\cdot Cap^{reg}_t\nonumber\\
    &\cdot \textit{Score}^{\textit{perf}}
    \label{eq:Income_reg1}\\
    \textit{Income}^{\textit{reg,perf}}  =\sum_{t=1}^{24}&\textit{Price}^{\textit{reg,perf}}_t\cdot Cap^{reg}_t\nonumber\\
    &\cdot R^{\textit{mileage}}\cdot \textit{Score}^{\textit{perf}}
    \label{eq:Income_reg}
\end{align}
\label{eq:Income_reg2}
\end{subequations}
\begin{subequations}
\begin{align}
    \textit{Mileage}^{RegA}&=\sum_{i=1}^n|RegA_i-RegA_{i-1}|\label{eq:Mileage1}\\
    \textit{Mileage}^{RegD}&=\sum_{i=1}^n|RegD_i-RegD_{i-1}|\\
    R^{mileage}&=\frac{\textit{Mileage}^{RegD}}{\textit{Mileage}^{RegA}}
    \label{eq:Mileage}
\end{align}  
\end{subequations}
$\textit{Price}^ \textit{reg,cap}_t$ is the day-ahead market capability clearing price while $\textit{Price}^\textit{{reg,perf}}_t$ is the day-ahead market performance clearing price. $Cap^{reg}_t$ is the hourly committed regulation capacity. $\textit{Score}^\textit{perf}$ is the historical performance score that reflects the accuracy of a regulation resource’s response to the regulation signal. The calculation of mileage ratio $R^{mileage}$ is the mileage of RegD $\textit{Mileage}^{RegD}$ divided by the mileage of RegA $\textit{Mileage}^{RegA}$, as described in formula \eqref{eq:Mileage1}-\eqref{eq:Mileage}\cite{PJMManual11}. The value of $R^{mileage}$ is approximately 3 so the battery storage following RegD earns three times the performance revenue. Additionally, RegD reduces the amount of battery storage's obligated reserved energy for it requires net zero energy over a 15-min period.

Battery storage is a favorable resource in the spinning reserve market since it can adjust its output power within seconds. The income is calculated as expressed in formula \eqref{eq:Income_res}, where $\textit{Price}^{res}$ is the day-ahead market spinning reserve clearing price and $ Cap^{res}$ is the hourly committed spinning reserve capacity.

\begin{equation}
\textit{Income}^{res}=\sum_{t=1}^{24}\textit{Price}^{res}_t\cdot Cap^{res}_t
    \label{eq:Income_res}
\end{equation}

\begin{figure}[htbp]
    \centering
    \includegraphics[width=0.95\linewidth]{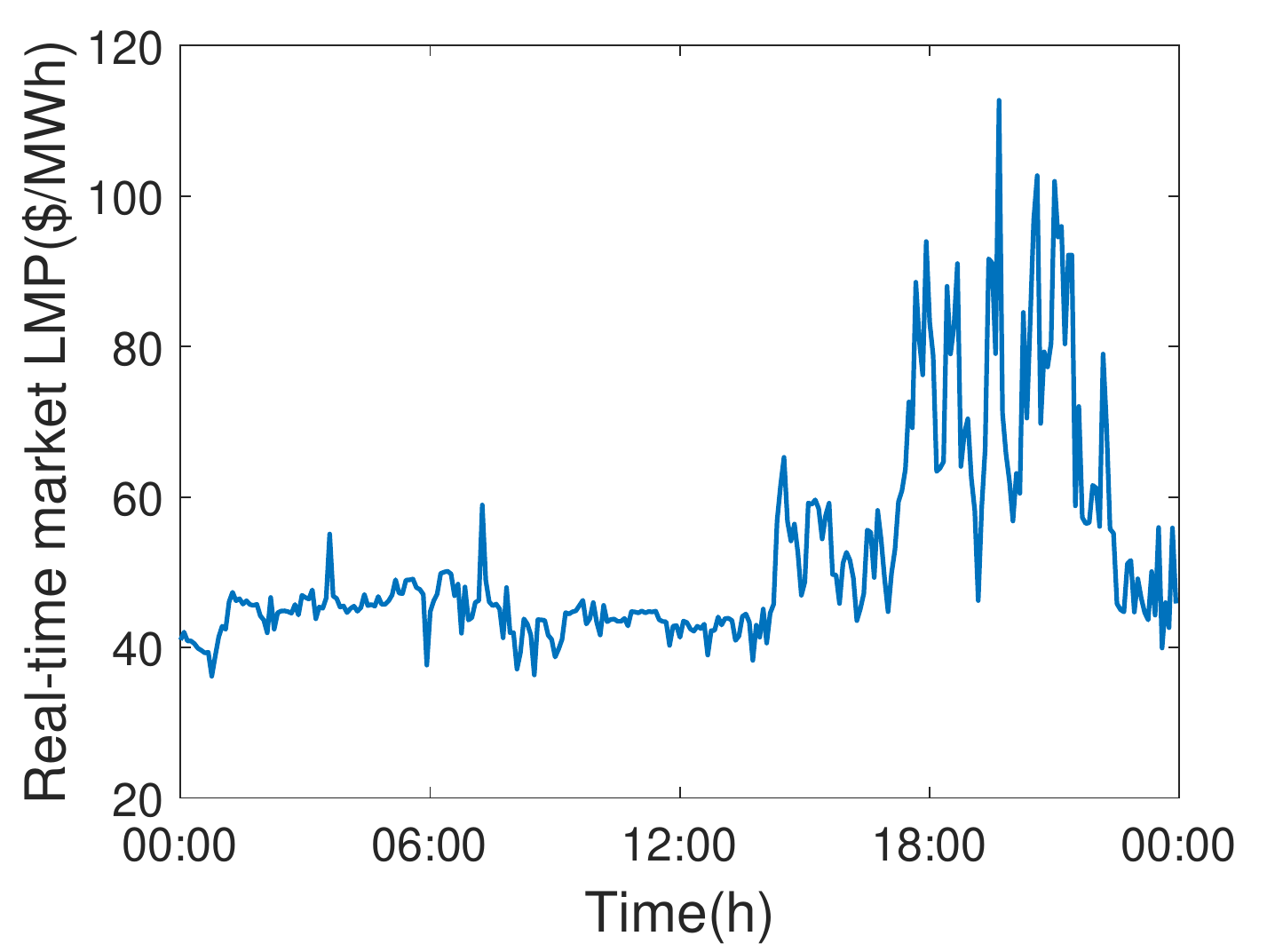}
    \caption{LMP price observations in PJM real-time market on May 1st, 2022}
    \label{fig:peak&valley}
\end{figure}

Load shifting in the energy market also benefits battery storage aside from the above-mentioned auxiliary services. Figure \ref{fig:peak&valley} shows the Locational Marginal Pricing(LMP) in PJM real-time market on May 1, 2022, in which the large price difference between peak and off-peak could render promising revenue to battery storage.

The calculation of load shifting income for battery storage is shown in formula \eqref{eq:Income_e1} under the constraint of formula \eqref{eq:Income_e}. 

\begin{subequations}
\begin{align}
    \!\textit{Income}^e \!&=\!\sum_{t=1}^{24}\textit{Price}^e_t\!\cdot \!\left( Cap^{ch}_t\!-Cap^{dch}_t\!\right) \label{eq:Income_e1}\\
    Cap^{ch}_i\cdot Cap^{dch}_i&=0,\quad i=1,2,\cdots,24
    \label{eq:Income_e}
\end{align}
\end{subequations}
$\textit{Price}^e_t$ refers to hourly LMP, $Cap^{ch}_t$ and $Cap^{dch}_t$ are the charging and discharging capacity within an hour respectively. It should be noted that the battery cannot be both charged and discharged in the same hour, which can be mathematically represented by formula \eqref{eq:Income_e}.

In the detailed consideration of battery degradation, this paper uses the concept of State of Health (SOH) to directly show the capacity degradation of battery storage, which describes the remaining capacity of the battery storage.

With the income of battery storage from ancillary service market as well as energy market included and the battery capacity degradation considered, this paper adopts the internal rate of return(IRR) to analyze the economic viability of whole life cycle for battery storage.

\section{Operation Simulation of Battery Storage Considering Capacity Degradation}\label{section3}

\subsection{Battery Degradation}\label{section3.1}
This paper adopts the optimal bidding strategy model considering the lifetime of battery storage. The capacity degradation of battery storage is considered in the operation simulation model referring to the comprehensive battery degradation model for NCM and LFP batteries in Olmos, 2021 \cite{olmos2021modelling}. 

The paper considers various battery degradation factors including full equivalent cycle $\textit{FEC}$, temperature $T$, depth of discharge $DoD$, charging rate $C_{ch}$, discharging rate $C_{dch}$, and average state of charge $mSOC$. The fitting formula based on Miner's rule is adopted in this paper, in which battery degradation is described as a function of full equivalent cycle, which is $\textit{f(FECs)}$. Other contributing factors are taken into account as the coefficients of $\textit{f(FECs)}$, as introduced in section \ref{section1}. Therefore, the capacity degradation model is described as equation \eqref{eq:SOH_ori}\cite{olmos2021modelling}. In equation \eqref{eq:SOH_ori}\, with $Q_{act}$ being the actual capacity of a lithium-ion battery and $Q_0$ defined as its initial capacity at beginning-of-life in nominal conditions, the SOH value is expressed with the initial capacity $Q_0$ defined as 100\% SOH. In order to simplify the modelling process, the influence of each stress factor is individualized and defined by a single expression, in which $f_{\lambda}$ refers to the function used to gather in a single equation the different expressions related to each stress factor. $SOH$ further expressed in details after complicated modelling and fitting, as shown in equation \eqref{eq:SOH_ori}, with the fitting parameters values of NCM and LFP batteries shown in table \ref{tab:degra para} \cite{olmos2021modelling}respectively.

\begin{table}[H]
\renewcommand{\arraystretch}{1.3}
\centering
\caption{Parameters of NCM \& LFP battery degradation model}
\label{tab:degra para}
    \begin{tabular}{p{2cm}|p{2cm}|p{2cm}}
    \toprule
    Parameter & NCM Value & LFP Value\\ 
    \midrule
    $\beta$ & 0.001673 & 0.003414\\
    $k_T$ & 21.6745 & 5.8755\\
    $k_{DoD}$ & 0.022 & -0.0046\\
    $k_{C_{ch}}$ & 0.2533 & 0.1038\\
    $k_{C_{dch}}$ & 0.1571 & 0.296\\
    $k_{mSOC}$ & -0.0212 & 0.0513\\
    $\alpha_{opt}$ & 0.915 & 0.869\\
    $T_{ref}$ & 293K & 293K\\
    $mSOC_{ref}$ & 42\% & 42\%\\
    \bottomrule
    \end{tabular}
\end{table}

\begin{equation}
    \begin{aligned}
    SOH =&\frac{Q_{act}}{Q_0}\cdot 100\\
    =&100-\Delta \textit{SOH}\\
    \!=&100\!-\!f_{\!\lambda}\!\cdot \!f\!_1(\textit{FEC})\!\cdot \!f\!_2\!(T\!)\!\cdot \!f\!_3(DoD)\!\cdot \!f\!_4(C_{ch})\\
    &\!\cdot \!f\!_5\!(C_{dch}\!)\!\cdot \!f\!_6\!(mSOC\!)\\
    =&100\!-\!\beta \!\cdot \exp(k_T\!\cdot \frac{T\!-T_{ref}}{T}\!+k_{DoD}\!\cdot DoD\!+k_{ch}\!\cdot C_{ch}\\
    &\!+k_{dch}\!\cdot C_{dch})\!\cdot\![1\!+\!b_{mSOC}\!\cdot mSOC\!\cdot\!(1\!-\!\frac{mSOC}{2\!\cdot\!mSOC_{\!ref}}\!)]\\
    &\!\cdot \textit{FEC}^{\alpha_{opt}}
    \end{aligned}
    \label{eq:SOH_ori}
\end{equation}

However, this comprehensive model causes very complex calculation. Therefore, this paper proposes the following reasonable simplifications of the calculation method.

\begin{itemize}
\item[1)]Depth of discharge $DoD$: Considering that the DoD of each cycle may not reach 100\%, this paper rewrites FEC (i.e. the number of cycles when DoD is 100\%) as the value of a specific DoD $d$ between 0-100\% and the number of cycles under this DoD $n_d$ according to the Miner's Rule, as described in formula \eqref{eq:FEC rewrite}.

\begin{equation}
    100\%\cdot \textit{FEC} = n_d\cdot d
    \label{eq:FEC rewrite}
\end{equation}

\item[2)]Charging rate $C_{ch}$ and discharging rate $C_{dch}$: At present, large-scale battery storage projects on renewable energy generation and the grid side have the characteristic of long battery storage duration, in which the charge-discharge rate will be limited to less than 1C. Under this constrained charge-discharge rate, laboratory results show that the effects of the spatial variations of local concentration and potential in the electrolyte are negligible.\cite{li2019development} Considering that this paper focus on the application in grid-scale battery storage, the impact of charge-discharge rate on battery degradation is relatively small. Therefore, this paper simplifies the charge-discharge rate as a constant value of 1C or 0.5C at the battery storage duration of 1h or 2h respectively. \cite{xu2017factoring}

\item[3)]Temperature T: Generally speaking, the temperature of the battery will be controlled under a reasonable range by the battery management system (BMS), and it is noted that the impact of temperature variation can be ignored before it reaches 40-50$^{\circ}$C \cite{olmos2021modelling}. Besides, considering the research purpose and the method of this paper, it is difficult to accurately simulate the temperature of battery storage. For the reasons mentioned above, the temperature is reasonably treated as a constant average value of 30$^{\circ}$C.

\item[4)]Mean State of Charge $mSOC$: For NCM and LFP batteries, 50\% is a empirical value of the mean State of Charge(mSOC).\cite{olmos2021modelling} Besides, since the relationship between mSOC and the degradation rate is a parabolic curve with its minimum point set between 40\%-50\% mSOC, the impact of mSOC on battery degradation is relatively small when mSOC changes around 50\%.\cite{olmos2021modelling} Therefore, This paper uses the mean value of 50\% for mSOC to simplify it.
\end{itemize}
After the simplifications above, the original degradation model is rewritten as formula \eqref{eq:SOH_rewrite}. The fitting parameters of NCM and LFP batteries are shown in table \ref{tab:degra para}.

\begin{equation}
\begin{aligned}
     SOH =&\frac{Q_{act}}{Q_0}\cdot 100 = 100-\Delta SOH\\
     =&100-\beta \cdot k_T\cdot k_{C_{ch}}\cdot k_{C_{dch}}\cdot k_{mSOC}\\&\cdot f_1(n_d\cdot d)\cdot f_3(d)
        \label{eq:SOH_rewrite}
\end{aligned}
\end{equation}

The number of cycles under a specific DoD $n_d$ could be rewritten to $N_d$, which means the maximum number of cycles under the DoD $d$. Since it is generally accepted that the battery should be replaced when its capacity decays to 80\% of the initial capacity, this paper set SOH to 80\% to obtain the relationship between $N_d$ and $d$ \cite{reniers2021unlocking}. Therefore,$N_d$ can be mathematically expressed by $d$. To more easily integrate this battery degradation model into the optimization for battery operation simulation model, this paper further transforms $N_d$ into $N_{100\%}$ through formula \eqref{eq:f(d)}, in which $\frac{1}{N_{100\%}}$ expresses the capacity degradation of one cycle when DoD $d$ is 100\%. After the simplifications and transformation above, the state of health(SOH) of the battery only depends on DoD $d$.

\begin{equation}
    \frac{1}{N_d}=f(d)=\frac{d^\alpha}{N_{100\%}}
    \label{eq:f(d)}
\end{equation}

For the calculation of DoD, this paper refers to the method of literature \cite{he2015optimal}, where $d_k$ is denoted as the DoD of the $k^{th}$ half cycle.

\subsection{Battery Operation Simulation}\label{}

Considering the complex clearing process of the electricity market, this paper ideally simplifies that the battery storage we are studying will 100\% be a winning bidder. Based on the optimal bidding strategy model, this paper proposes an operation simulation model of battery storage considering degradation.

The decision variables are $\textit{Cap}^{ch}_t$, $\textit{Cap}^{dch}_t$, $\textit{Cap}^{reg}_t$ and $\textit{Cap}^{res}_t$, which refer to the optimal capacity bids of charging, discharging, frequency regulation and spinning reserve for each hour in the next day (MW) according to clearing price of the corresponding electricity market. Since it is assumed that battery storage will win the bid unconditionally, the bidding capacity is the operation simulation capacity of the battery storage.

The bidding model is an income maximizing problem, in which battery degradation is treated as a part of the cost to avoid overly aggressive bidding. The objective function is described as formula \eqref{eq:obj}.

\begin{equation}
\begin{aligned}
    \textit{Objective}  =& \textit{Income}^e + \textit{Income}^{reg} + \textit{Income}^{res}\\
    & - \textit{Cost}^{op} - \textit{Cost}^{loss}
    \label{eq:obj}
\end{aligned}
\end{equation}

The revenue for load shifting in the energy market $\textit{Income}^e$ is calculated as formula \eqref{eq:rev_e}, where $\textit{Price}^e_t$ is the given LMP in the day-ahead market (\$/ MWh). Therefore, $\textit{Income}^e$ is described as a function of $\textit{Cap}^{ch}_t$ and $\textit{Cap}^{dch}_t$.

\begin{equation}
    \textit{Income}^e = \sum_{t=1}^{t=24} \textit{Price}^e_t\cdot (Cap^{dch}_t - Cap^{ch}_t)\cdot \Delta t
    \label{eq:rev_e}
\end{equation}

Formula \eqref{eq:rev_reg} calculates the revenue for frequency regulation $\textit{Income}^{reg}$, where $\textit{Income}^{reg,cap}$ and $\textit{Income}^{reg,perf}$ refer to capacity payment and performance payment respectively as described in formula \eqref{eq:Income_reg1}-\eqref{eq:Income_reg} in section \ref{section2}. Therefore, $\textit{Income}^{reg}$ is expressed as a function of $ \textit{Cap}^{reg}_t$.

\begin{equation}
\begin{aligned}
    &\textit{Income}^{reg} = \textit{Income}^{reg,cap} + \textit{Income}^{reg,perf}
    \label{eq:rev_reg}
\end{aligned}
\end{equation}

Battery storage's revenue for spinning reserve $\textit{Income}^{res}$ includes the auxiliary service part and the discharging income obtained in the energy market, as shown in the formula \eqref{eq:rev_res}. $\textit{Price}^{res}_t$ and $\textit{Price}^{e}_t$ are the clearing price of spinning reserve and LMP in the day-ahead market respectively, while $\textit{Prob}^{res}$ is the probability of spinning reserve's deployment. Therefore, $\textit{Income}^{res}$ can be calculated based on the value of $\textit{Cap}^{res}_t$.

\begin{equation}
    \textit{Income}^{res} = \sum_{t=1}^{t=24}( \textit{Price}^{res}_t\cdot Cap^{res}_t + \textit{Price}^e_t\cdot Cap^{res}_t\cdot \textit{Prob}^{res})
    \label{eq:rev_res}
\end{equation}

The operation and maintenance cost of battery storage $\textit{Cost}^{op}$ is shown in formula \eqref{eq:cost_op}, which includes the fixed part and the variable part. The fixed part $\textit{Cost}^\textit{op,fix}$ is determined by the rated power of battery storage $P_r$, while the variable part $\textit{Cost}^\textit{op,var}$ depends on the absolute value of the charge or discharge energy. The parameter $e^{reg}$ is the average hourly charge or discharge energy per MW frequency regulation capacity, which can be calculated from the RegD signal. $k_{\textit{fix}}$, $k_{var}$ are the empirical unit values of $\textit{Cost}^\textit{op,fix}$ and $\textit{Cost}^\textit{op,var}$ respectively. Therefore, $\textit{Cost}^\textit{op}$ is described by decision variables $\textit{Cap}^{ch}_t$, $\textit{Cap}^{dch}_t$, $ \textit{Cap}^{reg}_t$ and $\textit{Cap}^{res}_t$.

\begin{subequations}
\begin{align}
    \textit{Cost}^{op} =& \textit{Cost}^{op,fix} + \textit{Cost}^{op,var}\\
    \textit{Cost}^{op,fix} =& k_{fix}\cdot P_{r}\\
    \textit{Cost}^{op,var} =& \sum_{t=1}^{t=24} k_{var}\cdot (Cap^{dch}_t + Cap^{ch}_t \nonumber\\
    & \qquad+\textit{Prob}^{res}\cdot Cap^{res}_t \nonumber\\&\qquad+ 2\cdot e^{reg}\cdot Cap^{reg}_t)\cdot \Delta t
\end{align}
\label{eq:cost_op}
\end{subequations}

The degradation cost $ \textit{Cost}^{loss}$ is shown in formula \eqref{eq:cost_loss}. $\textit{Cost}^{bat}$ identifies the battery replacement cost, while the battery capacity loss $Cap^{loss}$ is calculated by formula \eqref{eq:cap_loss}. $N_{100\%}$ denotes the cycle number when DoD is 100\%. It should be noted that the DoD $d_k$ of the $k^{th}$ half cycle is calculated in this paper in reference of the method described in reference~\cite{he2015optimal}, rather than using the rainflow counting algorithm to determine the DoD $d$ of every complete cycle. Since it is assumed that the net energy within a day is zero, i.e. the final state of battery energy is the same as the initial state in each day, and each charging half-cycle corresponds to a discharging half-cycle, it is reasonable to adopt this method of half cycle.

\begin{align}
    \textit{Cost}^{loss}&=\textit{Cost}^{bat}\cdot Cap^{loss} \label{eq:cost_loss} \\
    \textit{Cap}^{loss}&=\sum_{k\in C}\frac{1}{2}\cdot \frac{d_k^\alpha}{N_{100\%}} \label{eq:cap_loss}
\end{align}

In this way, $\textit{Cost}^{loss}$ is determined only by DoD without mathematical relationship with the decision variables. Therefore, the decision variables are decoupled from the calculation of DoD for each charging-discharging cycle of the battery, which allows for better integration of the battery degradation model with the battery operation simulation.

 \eqref{eq:constr_1}-\eqref{eq:constr_2} model the operational constraints of the battery storage.

\textit{1) Capacity Constraints:}

The capacity bids of charging, discharging, frequency regulation, and spinning reserve must be kept between zero and their upper limits respectively. The binary variable $b_t$ denotes that the battery charges when $b_t$ is 1 while discharges when $b_t$ is 0. 

\begin{subequations}
    \begin{align}
        & 0 \leq Cap^{dch}_t\cdot b_t \leq P_r\\    
        & 0 \leq Cap^{ch}_t\cdot (1-b_t) \leq P_r\\
        & 0 \leq Cap^{reg}_t\leq P_r\\   
        & 0 \leq Cap^{res}_t\leq P_r
    \end{align}
\label{eq:constr_1}
\end{subequations}
The sum of the capacity bids should also be kept below its upper limits. In discharging half cycles, the output power provided must correspond to $Cap^{res}_t$ of the battery storage once its spinning reserve is called.

\begin{subequations}
    \begin{align}
         Cap^{dch}_t + Cap^{reg}_t + Cap^{res}_t &\leq P_r\\
         Cap^{ch}_t + Cap^{reg}_t &\leq P_r
    \end{align}
\end{subequations}

\textit{2) Energy constraints:}

The energy held in battery storage is kept within its upper and lower limits, where $e_0$ is the initial SoC and $e_t$ identifies the SoC at hour $t$ ($t$=1,2,\dots,24).

\begin{equation}
    E_{min} \leq e_t \leq E_{max}, \qquad t=0,1,\cdots,24
\label{eq:E_max}
\end{equation}

A battery must be able to maintain the fully-deployed output level for at least a specific time interval according to the market, which is 1 h for spinning reserve service and 15 min for regulation service in this paper\cite{he2015optimal}, described by $t^{res}$ and $t^{reg}$ respectively. $\textit{eff}_{ch}$ and $\textit{eff}_{dch}$ denote the charging and discharging efficiency of battery.

\begin{subequations}
    \begin{align}
     \frac{Cap^{dch}_t\cdot \Delta t + Cap^{res}_t\cdot t^{res} + Cap^{reg}_t\cdot t^{reg}}{\textit{eff}_{dch}} &\leq e_t\\
     e_t + (Cap^{ch}_t\cdot \Delta t + Cap^{reg}_t\cdot t^{reg})\cdot \textit{eff}_{ch} &\leq E_{max}
    \end{align}
\end{subequations}

\textit{3) State of Charge Constraints:}

$\Delta e_t$ represents the amount of energy change in hour $t$ due to energy selling and purchasing, reserve deployment, energy loss in providing regulation service, and self-discharge.

\begin{equation}
    \begin{aligned}
    \Delta e_t =& e_t - e_{t-1}\cdot r_{\textit{self}}\\
    \Delta e_t =& \left(\right.Cap^{ch}_t\cdot \textit{eff}_{ch} - \frac{Cap^{dch}_t + Cap^{res}_t\cdot Prob^{res}}{\textit{eff}_{dch}}\\
    &-\left(\frac{e^{reg}\cdot Cap^{reg}_t}{\textit{eff}_{dch}} - e^{reg}\cdot Cap^{reg}_t\cdot \textit{eff}_{ch}\right)\left.\right) \cdot \Delta t
    \end{aligned}
\end{equation}

Since battery storage mainly profits through frequency regulation, the initial SoC of each day would not cause much difference to the revenue of battery storage. Therefore, for the convenience of calculation, it is assumed that the battery returns to the initial SoC after a day's service, which is also the minimum value of the day.

\begin{equation}
    \begin{aligned}
    & e_0 = e_{24}= E_{min}
    \end{aligned}
     \label{eq:constr_2}
\end{equation}

In conclusion, the operation simulation of battery storage considering capacity degradation is modeled as follow:

\begin{alignat*}{2}
& \underset{\tiny\begin{aligned}&\textit{Cap}^{ch}_t, \textit{Cap}^{dch}_t, \\& \textit{Cap}^{reg}_t, \textit{Cap}^{res}_t\end{aligned}}{\text{maximize}} &\quad& \textit{formula}~\eqref{eq:obj} \\
&&& \text{(substituting formulas}~\eqref{eq:Income_reg2},~\eqref{eq:rev_e}\text{,} \\
&&&\text{and}~\eqref{eq:cap_loss}~\text{into formula}~\eqref{eq:obj})\\
& \text{subject to} &\quad& \textit{formula}~\eqref{eq:constr_1}-\eqref{eq:constr_2}
\end{alignat*}

Mixed-integer programming(MIP) optimization technique is adopted, employing the yalmip+gurobi toolbox.

\section{Economic Viability Analysis}\label{section4}

\subsection{Life Cycle Revenue and Cost Structure}\label{section4.1}

The life cycle revenue and cost of battery storage includes the variable part and the fixed part. The variable part includes the revenue of load shifting, regulation and reserve, as well as the corresponding operation and maintenance cost and battery degradation cost, which can be calculated through the model proposed in section \ref{section3}. The fixed part covers other one-time cost such as battery purchasing cost, power station design and construction cost, and battery recycling income at the end of the investment period.

\begin{figure}[htbp]
    \centering   \includegraphics[width=1.0\linewidth]{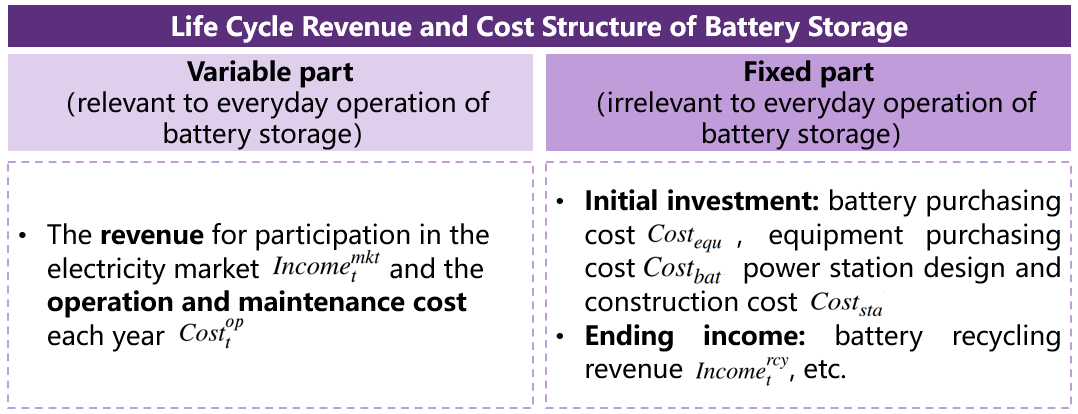}
    \caption{Life cycle revenue and cost structure of battery storage}
    \label{fig:structure}
\end{figure}
\subsection{Revenue and Cost Calculation}\label{section4.2}
As described in section \ref{section3}, the daily operation of battery storage is obtained through the battery storage operation simulation model proposed in this paper, in which battery storage decides its bidding capacity for different markets $\textit{Cap}^{ch}_t$, $\textit{Cap}^{dch}_t$, $\textit{Cap}^{reg}_t$ and $\textit{Cap}^{res}_t$ under optimal bidding strategy.By solving this optimize problem, the optimal capacity bids of charging, discharging, frequency regulation and spinning reserve for each hour in the next day are uniquely determined by the market price of energy market and auxiliary service market $\textit{price}^e_i$, $\textit{price}^{reg}_i$, $\textit{price}^{res}_i$. Therefore, the revenue $\mathit{income}_i$, cost $\mathit{cost}^{op}_i$, and degradation of battery storage $\mathit{cap}^{loss}_i$ can be calculated, in which the battery storage gross income of day $i$ $\mathit{income}_i$ is the sum of $\textit{Income}^e$, $\textit{Income}^{reg}$, and $\textit{Income}^{res}$. Generally speaking, the model can be abstracted as a schematic formula \eqref{eq:MODEL}. The subscript $i$ represents the date in the set of selected operation simulation dates.

\begin{equation}
\label{eq:MODEL}
\begin{aligned}
    & \left(\mathit{income}_i, \mathit{cost}^{op}_i, \mathit{cap}^{loss}_i\right)\\
    & =\mathit{Model}\left(\mathit{price}^e_i, \mathit{price}^{reg}_i, \mathit{price}^{res}_i\right)
\end{aligned}
\end{equation}

This paper uses the operation simulation model to calculate the daily operation of the battery storage and obtains the corresponding daily battery capacity loss. The life of the battery storage ends when the capacity fading percentage is accumulated to 20\%, 

Since the life of battery storage generally reaches 8-15 years, we need to conduct operation simulation of the data in each day of 15 years. Considering its huge workload, this paper selects typical days in each year within the life of the battery storage to simplifies calculation. Specifically, we adopt two different methods to select typical days: picking out 36 days in each year considering time factors such as date and holiday, as well as a clustering method.

In the selection of 36 typical days in each year, this paper fully considers the time factors that may bring about changes in the operation of the electricity market, such as months, weeks, and festivals. In terms of specific choices, the three days is chosen in the first, middle, and last ten days of each month respectively, of which two days are weekdays while the other day is selected from weekends. Since this paper uses the electricity market data of US, considering the factor of festival in the US, the Independence Day on July 4, Halloween on November 1, and Christmas on December 25 are also added to the 36 typical days of the year. After obtaining the real electricity market price data on typical days of each year, this paper re-select the date with much more price deviations from the normal level has been re-selected to ensure the reasonableness of final linear extension of typical days.

\begin{figure}[htbp]
    \centering
    \includegraphics[width=0.85\linewidth]{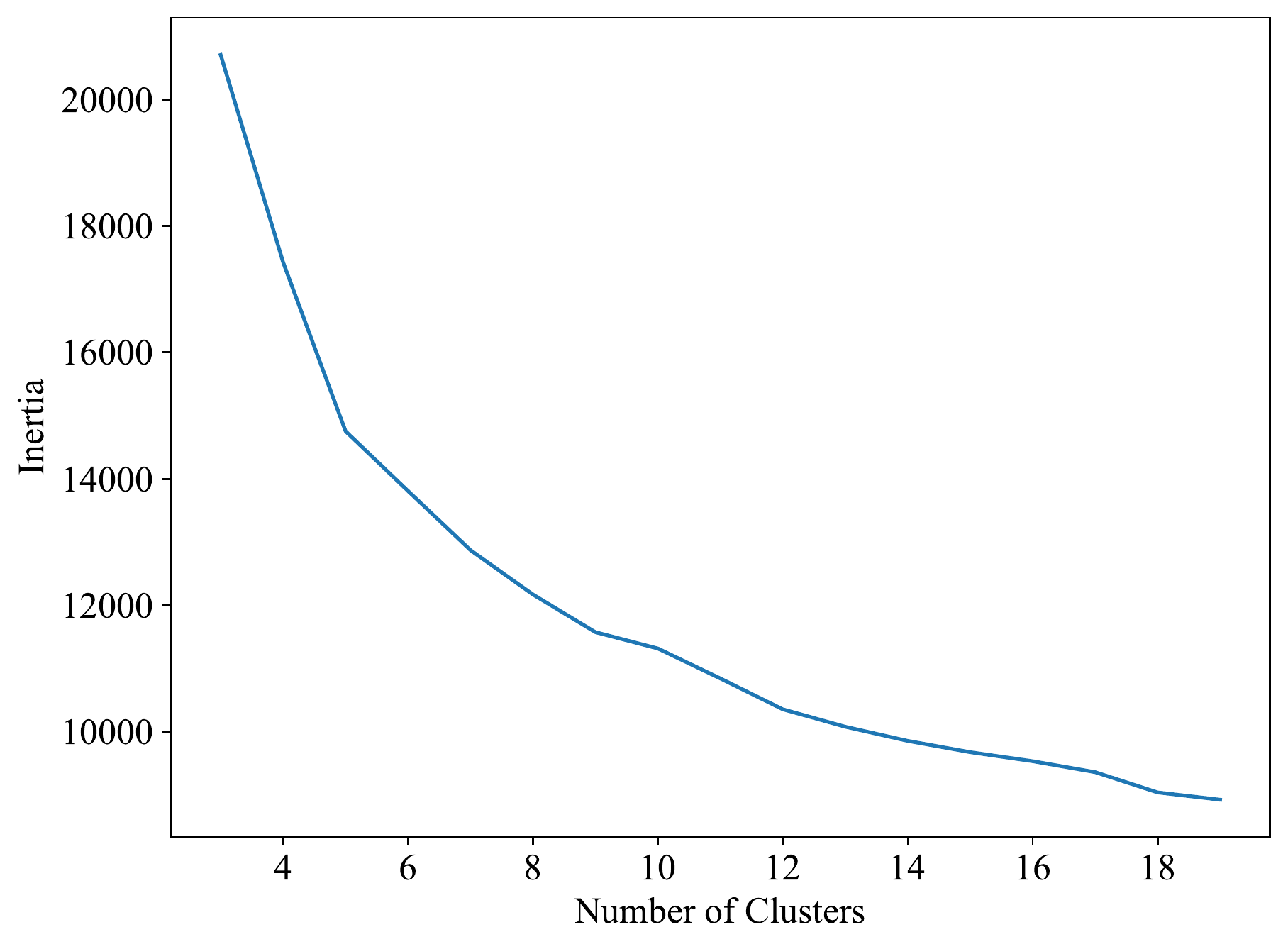}
    \caption{The relationship between cluster number and inertia(within-cluster sum-of-squares criterion)}
    \label{fig:n_cluster}
\end{figure}

\begin{figure*}[htbp]
    \centering
    \includegraphics[width=1\linewidth]{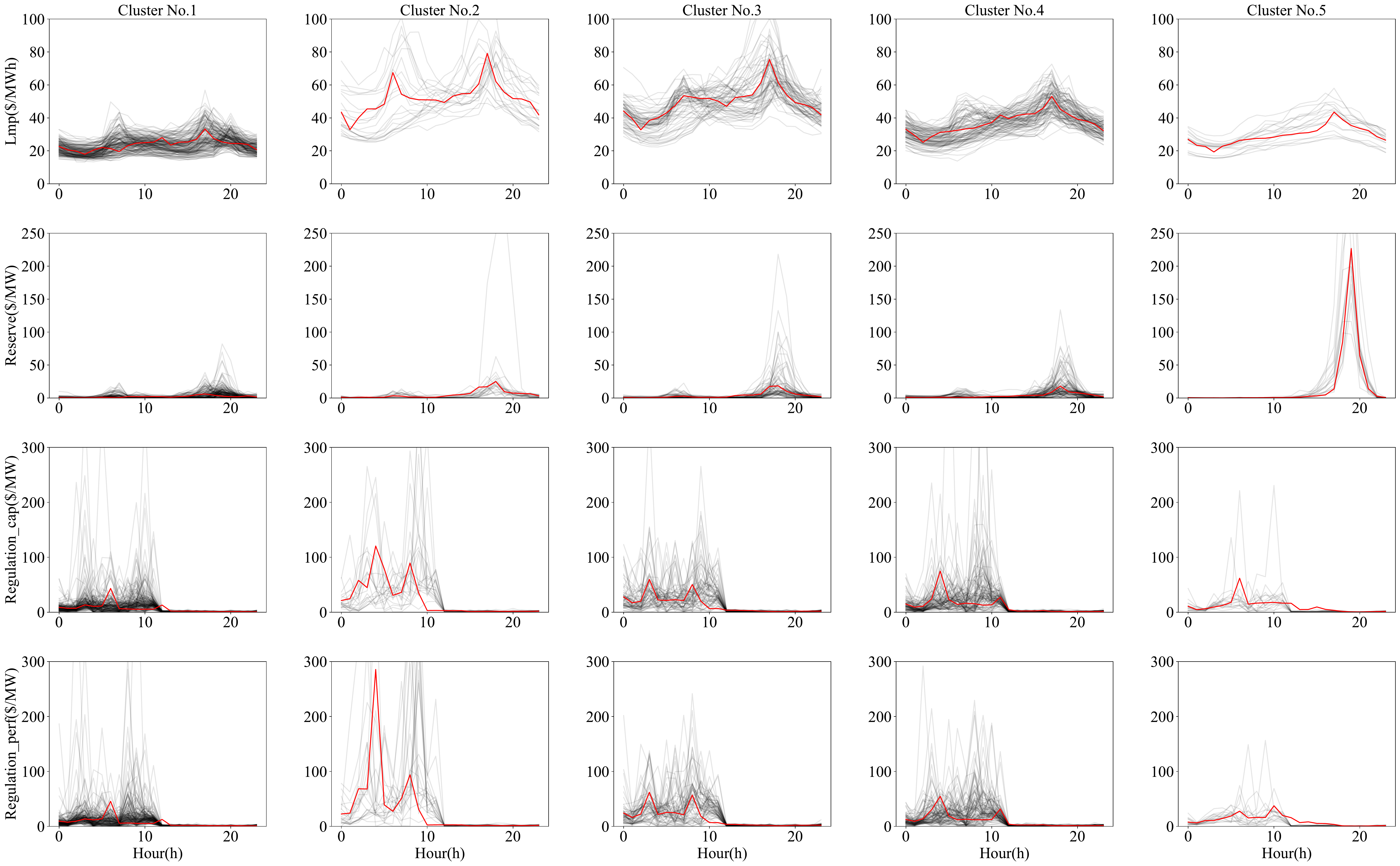}
    \caption{Clustering result of electricity market price when $n_{cluster}$=5}
    \label{fig:clustering}
\end{figure*}

According to the operation simulation model and equation \eqref{eq:MODEL}, this paper calculates battery storage gross income $\textit{Income}_T$ in year $T$, operation and maintenance cost $\textit{Cost}^{op}_T$, and capacity loss $\textit{Cap}^{loss}_T$ by the end of year $T$, the increase of which are ten times that of the typical days in year $T$,  as calculated in equation \eqref{eq:income_36}. $\textit{income}_{T,i}$, $\textit{cost}^{op}_{T,i}$, and $\textit{cap}^{loss}_{T,i}$ indicates the data of typical day $i$ in year $T$.

\begin{subequations}
\begin{align}
\textit{Income}_T&=\textit{Income}_{T-1}+10\cdot \sum_{i=1}^{i=36}\textit{income}_{T,i}\\ \textit{Cost}^{op}_T&=\textit{Cost}^{op}_{T-1}+10\cdot \sum_{i=1}^{i=36}\textit{cost}^{op}_{T,i}\\ \textit{Cap}^{loss}_T&=\textit{Cap}^{loss}_{T-1}+10\cdot \sum_{i=1}^{i=36}\textit{cap}^{loss}_{T,i}
\end{align}
\label{eq:income_36}
\end{subequations}
Apart from the method above, we further adopt the clustering method based on the electricity market price in 2021. We apply the Kmeans algorithm with dynamic time warping metric, and the implementation by tslearn\cite{JMLR:v21:20-091} is adopted. The inertia of the result, which is the within-cluster sum-of-squares criterion, decreases as the number of clusters $n_{cluster}$ increases, signifying higher accuracy. It can be analyzed from figure \ref{fig:n_cluster} that the first inflection point appears when $n_{cluster}$ is 5. Therefore, the data is clustered into 5 types for the price of load shifting, spinning reserve, and frequency regulation respectively balancing calculation accuracy and simplicity, as shown in figure \ref{fig:clustering}.

According to the operation simulation model and equation \eqref{eq:MODEL}, battery storage gross income of the year $\textit{Income}_T$, operation and maintenance cost $\textit{Cost}^{op}_T$, and capacity loss $\textit{Cap}^{loss}_T$ are the sum of n clusters of the days by the end of year $T$, as calculated in equation \eqref{eq:income_cluster}. $\mathit{D_{T,i}}$ represents the number of days in cluster $i$ in year $T$, while $\textit{income}_i$, $\textit{cost}^{op}_i$, and $\textit{cap}^{loss}_i$ indicates the data of typical days in cluster $i$.

\begin{subequations}
\begin{align}
\textit{Income}_T&=\textit{Income}_{T-1}+\sum_{i=1}^{i=n}\mathit{D_{T,i}}\cdot\textit{income}_{T,i}\\ 
\textit{Cost}^{op}_T&=\textit{Cost}^{op}_{T-1}+\sum_{i=1}^{i=n}\mathit{D_{T,i}}\cdot \textit{cost}^{op}_{T,i}\\ \textit{Cap}^{loss}_T&=\textit{Cap}^{loss}_{T-1}+\sum_{i=1}^{i=n}\mathit{D_{T,i}}\cdot \textit{cap}^{loss}_{T,i}    
\end{align}
\label{eq:income_cluster}
\end{subequations}
The detailed calculation method is shown as in algorithm \ref{alg:MODEL} and \ref{alg:MODEL2}.

\begin{algorithm}
  \caption{Economic viability analysis of battery storage using the method of 36 typical days}
  \label{alg:MODEL}
  \begin{algorithmic}[1]
  \REQUIRE set of typical days $\textit{M}$, set of electricity market price $\textit{Price}^e, \textit{Price}^{reg}, \textit{Price}^{res}$.
  \STATE Initializes battery storage gross income $\textit{Income}$, operation and maintenance cost $\textit{Cost}^{op}$, capacity loss $\textit{Cap}^{loss}$
  \WHILE{accumulated capacity loss $\textit{Cap}^{loss}<20\%$}
  \STATE The battery storage project income, operation and maintenance cost and capacity loss of the year can be calculated according to the operation simulation model and equation \eqref{eq:income_36}.
    \ENDWHILE
  \end{algorithmic}
\end{algorithm}

\begin{algorithm}
  \caption{Economic viability analysis of battery storage using clustering method}
  \label{alg:MODEL2}
  \begin{algorithmic}[1]
    \REQUIRE number of clusters $n$, number of operation simulation days in $n$ clusters $\mathit{D_{T,1}}, \mathit{D_{T,2}}, \dots, \mathit{D_{T,n}}$ and electricity market price of cluster $i$, $\mathit{Price}^e_{T,i}, \mathit{Price}^{reg}_{T,i}, \mathit{Price}^{res}_{T,i}$ in each year $T$
    \STATE Initializes battery storage gross income $\textit{Income}$, operation and maintenance cost $\textit{Cost}^{op}$, capacity loss $\textit{Cap}^{loss}$
    \WHILE{accumulated capacity loss $\textit{Cap}^{loss}<20\%$}
    \STATE The battery storage project income, operation and maintenance cost and capacity loss of the year can be calculated according to the operation simulation model and equation \eqref{eq:income_cluster},
     \ENDWHILE
  \end{algorithmic}
\end{algorithm}
As for the fixed revenue and cost, battery purchasing cost takes up a significant portion of it,  which is usually predicted based on learning curve. The experienced rate of the falling battery purchasing cost is estimated on the base of Wright's Law\cite{schmidt2017future}\cite{hsieh2019learning}\cite{schmidt2019projecting}. Equipment purchasing cost as well as power station design and construction cost are also included in the early investment of battery storage projects. Additional revenue from battery recycling of energy storage at the end of its life is relatively low at present.

\subsection{Internal Investment Rate for Economic Viability}
The internal investment rate (IRR) of a specific project is the discount rate that makes the net present value (NPV) of it exactly zero. The investment can be profitable when IRR is greater than the actual discount rate, and the potential rate of return on the investment positively correlates with the difference between IRR and the actual discount rate. Therefore, IRR is adopted as a comprehensive indicator to analyse the life cycle economic viability of battery storage, whose calculation formula is shown as \eqref{eq:IRR}. $C_0$ is the initial investment cost, while $C_t$ is the net cash flow for year $t$.

\begin{equation}
    0=\textit{NPV}=\sum_{t=1}^n\frac{C_t}{(1+\textit{IRR})^t}-C_0
    \label{eq:IRR}
\end{equation}
As mentioned in section \ref{section4.2}, the variable revenue and cost includes the revenue of battery storage for participation in the electricity market each year $\textit{Income}_t^{\textit{mkt}}$ and the operation and maintenance cost each year $\textit{Cost}_t^{op}$, while the fixed part covers battery purchasing cost $\textit{Cost}_{\textit {bat}}$, equipment purchasing cost $\textit{Cost}_{\textit{equ}}$, power station design and construction cost $\textit{Cost}_{\textit{sta}}$ and battery recycling revenue $\textit{Income}_t^{\textit{rcy}}$. According to formula \eqref{eq:cash flow}, the cash flow of each year in the entire battery storage life cycle of $n$ years can be calculated.

\begin{subequations}
\begin{align}
    C_0&=-(\textit{Cost}_{\textit{bat}}+\textit{Cost}_{\textit{equ}}+\textit{Cost}_{\textit{sta}})\\
    C_t&=\textit{Income}_t^{\textit{mkt}}-\textit{Cost}_t^{op}, \qquad t=1,2,\cdots,n-1\\
    C_n&=\textit{Income}_t^{\textit{mkt}}-\textit{Cost}_t^{op}+\textit{Income}_t^{\textit{rcy}}
\end{align}
    \label{eq:cash flow}
\end{subequations}
Based on formula \eqref{eq:cash flow}, IRR can be obtained through formula \eqref{eq:IRR}.
\subsection{Overall Economic Viability Analysis Model}
The battery operation simulation model of day $i$ is a mixed-integer programming(MIP) model which intends to maximum formula~\eqref{eq:obj} with constraints of formula~\eqref{eq:constr_1}-\eqref{eq:constr_2} under the specific electricity market prices. The decision variables include the bidding capacity of battery storage for different markets $\textit{Cap}^{ch}_t$, $\textit{Cap}^{dch}_t$, $\textit{Cap}^{reg}_t$ and $\textit{Cap}^{res}_t$ of hour $t$. The objective function of formula~\eqref{eq:obj} is 
correlated with the decision variables through formulas~\eqref{eq:Income_reg2}, \eqref{eq:rev_e}-\eqref{eq:cap_loss}. After solving this optimization problem, the battery operation model of day $i$ can be abstracted as formula \eqref{eq:MODEL}, in which the battery storage gross income of day $i$ $\mathit{income}_i$ is the sum of $\textit{Income}^e$, $\textit{Income}^{reg}$, and $\textit{Income}^{res}$. Therefore, the incomes and costs of battery storage can be summed up through formula \eqref{eq:income_36} or formula \eqref{eq:income_cluster} according to the different methods adopted to obtain the total incomes and costs of battery storage by the end of year $T$. Hence, we can obtain the gross income of each year $t$, and further calculate the cash flow of each year through formula \eqref{eq:cash flow}. Based on the cash flow $C_t$ of each year, we can calculate $\textit{IRR}$ through formula \eqref{eq:IRR}.

\section{Case Study}
\subsection{Basic setup}
This paper adopts electricity market data in 2021 from PJM\
(Pennsylvania-New Jersey-Maryland Interconnection) and CAISO (California Independent System Operator). Day-ahead prices for energy and frequency regulation as well as RegD data are acquired from PJM while spinning reserve market data is gained from CAISO, in which day-ahead prices for energy selects the price of a node with a voltage level of 138kV in consideration of the grid-connected voltage level of a large-scale energy storage project. The reason we use the day-ahead spinning reserve price of CASIO rather than PJM is that PJM includes an extra capacity market which is beyond the scope of this paper, resulting in its relatively low price of spinning reserve. Considering that there are many factors affecting electricity prices, resulting in the low accuracy of long-term price forecasts, this paper assumes that the electricity market price data in future years are the same as the historical data in 2021. Since the RegD signal of 2021 is not collected, it is assumed to be the same as that of 2020.

In addition, considering the time difference between PJM and CAISO electricity market due to geographical location, the spinning reserve price data for the CAISO market has been shifted for 4 hours.


Since NCM and LFP are both competitive currently, these two battery technology routes are analyzed respectively.

The degradation model of LFP and NCM battery storage project can be formulated as \eqref{eq:fade_NMC} according to section \ref{section3.1}.

\begin{equation}
    f_{\textit{LFP}}(d)=\frac{d^\alpha}{N_{100\%}}=\frac{d}{13627}
    \label{eq:fade_LFP}
\end{equation}

\begin{equation}
    f_{\textit{NCM}}(d)=\frac{d^\alpha}{N_{100\%}}=\frac{d}{10420}
    \label{eq:fade_NMC}
\end{equation}
This paper refers to a grid-side LFP battery energy storage project provided by China Southern Power Grid to set relevant parameters as shown in table \ref{tab:stor_para}, including the battery self-discharge rate $r_\textit{self}$, the score of historical price $\textit{Score}^{\textit{perf}}$, and the mileage ratio $R^{\textit{mileage}}$\cite{he2015optimal}. 

\begin{table}[H]
\renewcommand{\arraystretch}{1.3}
\centering
\caption{Typical grid side energy storage project battery parameters}
\label{tab:stor_para}
    \begin{tabular}{p{0.85cm} p{0.85cm} p{0.65cm} p{0.65cm} p{0.65cm} p{1cm} p{1cm}}
    \toprule
    $P_r$ & $E_r$ & $\textit{eff}_{ch}$ & $\textit{eff}_{dch}$ & $r_{\textit{self}}$ & $\textit{Score}^{\textit{perf}}$ & $R^{\textit{mileage}}$\\ 
    \midrule
    10MW & 20MWh & 94\% & 94\% & 1\% & 0.9 & 2.8\\
    \bottomrule
    \end{tabular}
\end{table}

According to this project of China Southern Power Grid, the price parameters of the example LFP project can be calculated, as listed in table \ref{tab:stor_price}.

\begin{table}[H]
\renewcommand{\arraystretch}{1.3}
\centering
\caption{Typical grid-side LFP battery storage project price parameters}
\label{tab:stor_price}
    \begin{tabular}{p{2.5cm} p{2.5cm} p{2.5cm}}
    \toprule
    Battery purchasing cost & Equipment purchasing cost & Station design and construction cost\\
    $\textit{Cost}_{\textit{bat,pur}}$ & $\textit{Cost}_{equ}$ & $\textit{Cost}_{sta}$\\
    \midrule
    \$4,150,000 & \$1,950,000 & \$1,280,000\\
    \bottomrule
    \end{tabular}
\end{table}
Considering that the price of an NCM battery is 1.5-2 times that of an LFP battery due to the containment of precious metal, the battery purchase cost of NCM is assumed as 2 times of LFP, as shown in formula \eqref{eq:NMC2LFP}. The equipment purchase cost and the power station design and construction cost of LFP are NCM are assumed to be the same and constant over time, as expressed in the above table.

\begin{equation}
    \textit{Cost}_{\textit{bat,NMC}}=2\cdot \textit{Cost}_{\textit{bat,LFP}}
    \label{eq:NMC2LFP}
\end{equation}
The battery replacement cost is calculated through the battery price converted to the time when the battery is retired. The life cycle of the LFP and NCM battery example projects is calculated to be about 12-14 years, which corresponds to its retirement in 2033-2035. According to literature \cite{schmidt2019projecting}, which forecasted the lithium-ion battery cost reduction in the future with the price data of 2015 selected as a baseline, the lithium-ion battery cost in 2020/2025/2030/2035 would reduce to 55\%/34\%/23\%/18\%/ of the battery cost in 2015. Therefore, this paper reasonably assumes that the purchase cost of the battery will decrease after 12-14 years to $k_{dec}=\frac{4}{11}$ of that in 2021. The battery replacement cost $\textit{Cost}_{\textit{bat,exc}}$ is expressed as formula\eqref{eq:BatteryReplace}.

\begin{equation}
    \textit{Cost}_{\textit{bat,exc}}=k_{dec}\cdot \textit{Cost}_{\textit{bat,pur}}=\frac{4}{11}\cdot \textit{Cost}_{\textit{bat,pur}}
    \label{eq:BatteryReplace}
\end{equation}

The recycling income includes that of the battery and equipment.  Considering the impact of the maturity of battery recycling technology and the expansion of industrial scale in the future, the recycling value ratio of retired NCM/LFP batteries is assumed to be 10\%/30\%, while the equipment recycling value ratio is set as 40\%, as shown in formula \eqref{eq:recycle}.

\begin{subequations}
    \begin{align}
    & \textit{Income}_{\textit{rcy,NCM}}=10\% \cdot \textit{Cost}_{bat}+40\%\cdot \textit{Cost}_{equ}\\
    & \textit{Income}_{\textit{rcy,LFP}}=30\% \cdot \textit{Cost}_{bat}+40\%\cdot \textit{Cost}_{equ}
    \end{align}
    \label{eq:recycle}
\end{subequations}
In conclusion, the price parameters of the LFP and NCM battery storage project for the case study are summarized in the table \ref{tab:stor_price_LFP} and table \ref{tab:stor_price_NMC}.

\begin{table}[H]
\renewcommand{\arraystretch}{1.3}
\centering
\caption{Price parameters of the LFP battery storage example project}
\label{tab:stor_price_LFP}
    \begin{tabular}{p{1.5cm} p{1.5cm} p{1.5cm} p{1.5cm} p{2cm} p{1.5cm} p{1.5cm}}
    \toprule
    $\textit{Cost}_{\textit{bat,pur}}$ & $\textit{Cost}_{\textit{bat,exc}}$ & $\textit{Cost}_{equ}$ & $\textit{Cost}_{sta}$\\
    \midrule
    \$4,150k & \$1,510k & \$1,950k & \$1,280k \\
    \midrule
    \multicolumn{2}{c}{$\textit{Income}_{rcy}$} & $k_{\textit{fix}}$ & $k_{\textit{var}}$\\
    \midrule
    \multicolumn{2}{c}{\$1,190k} & \$10/MW & \$0.5/MWh\\
    \bottomrule
    \end{tabular}
\end{table}

\begin{table}[H]
\renewcommand{\arraystretch}{1.3}
\centering
\caption{Price parameters of the NCM battery storage example project}
\label{tab:stor_price_NMC}
    \begin{tabular}{p{1.5cm} p{1.5cm} p{1.5cm} p{1.5cm} p{2cm} p{1.5cm} p{1.5cm}}
    \toprule
    $\textit{Cost}_{\textit{bat,pur}}$ & $\textit{Cost}_{\textit{bat,exc}}$ & $\textit{Cost}_{equ}$ & $\textit{Cost}_{sta}$ \\
    \midrule
    \$8,330k & \$3,020k & \$1,950k & \$1,280k \\
    \midrule
    \multicolumn{2}{c}{$\textit{Income}_{rcy}$} & $k_{\textit{fix}}$ & $k_{\textit{var}}$\\
    \midrule
    \multicolumn{2}{c}{\$3,270k} & \$10/MW & \$0.5/MWh\\
    \bottomrule
    \end{tabular}
\end{table}

\subsection{Example Result of Operation Simulation Model}

\begin{figure}[htbp]
  \centering
  \includegraphics[width=1.0\linewidth]{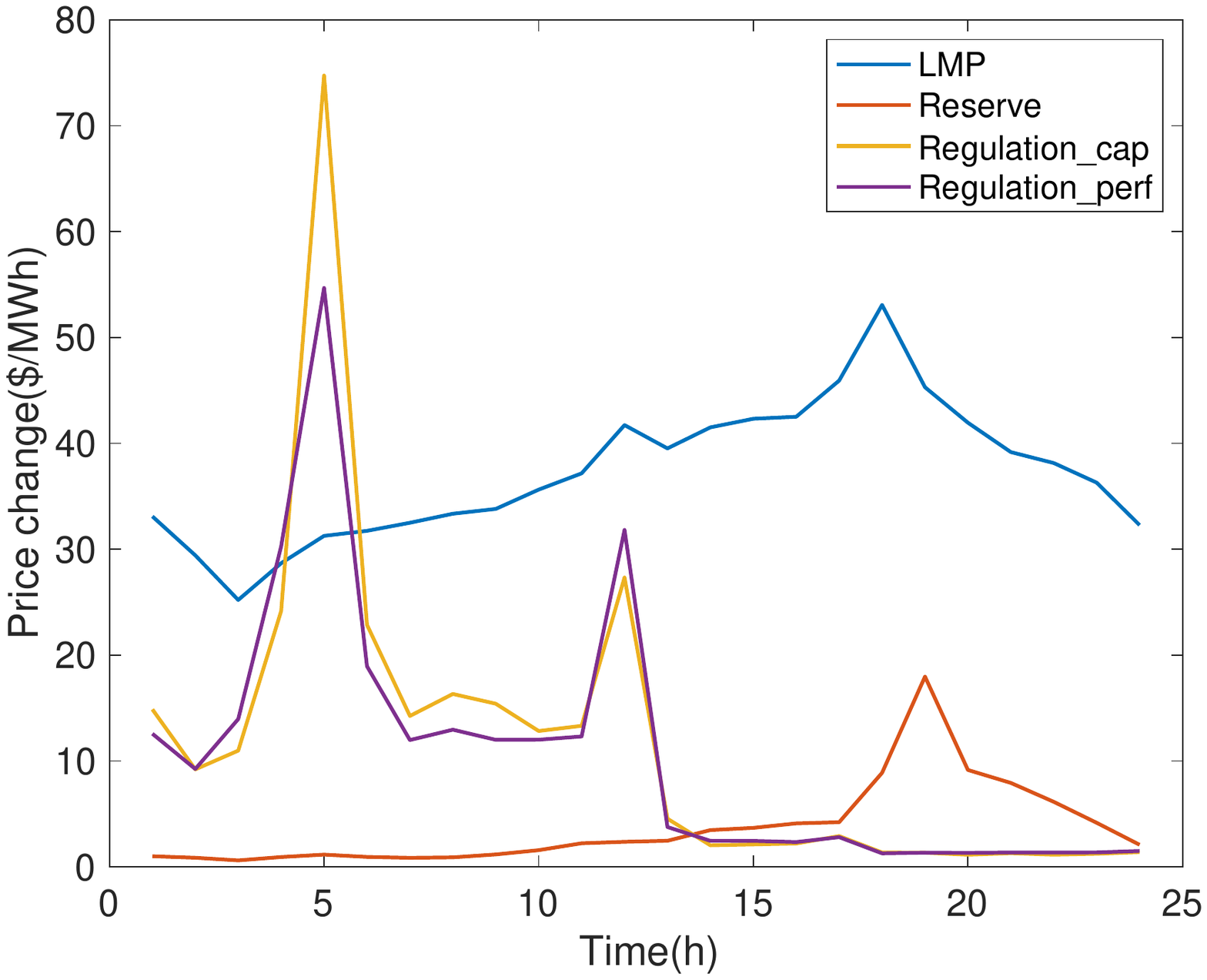}
  \caption{The price data of PJM and CASIO electricity market in 2021 of cluster 4}
  \label{fig:cluster4_price}
\end{figure}

To express the operation simulation result in detail, this section takes the operation simulation result of the days in cluster 4 as an example to show the solution of the LFP battery storage operation simulation model. Figure \ref{fig:cluster4_price} expresses the electricity market price of cluster 4.

\begin{figure}[htbp]
  \centering
  \includegraphics[width=1.0\linewidth]{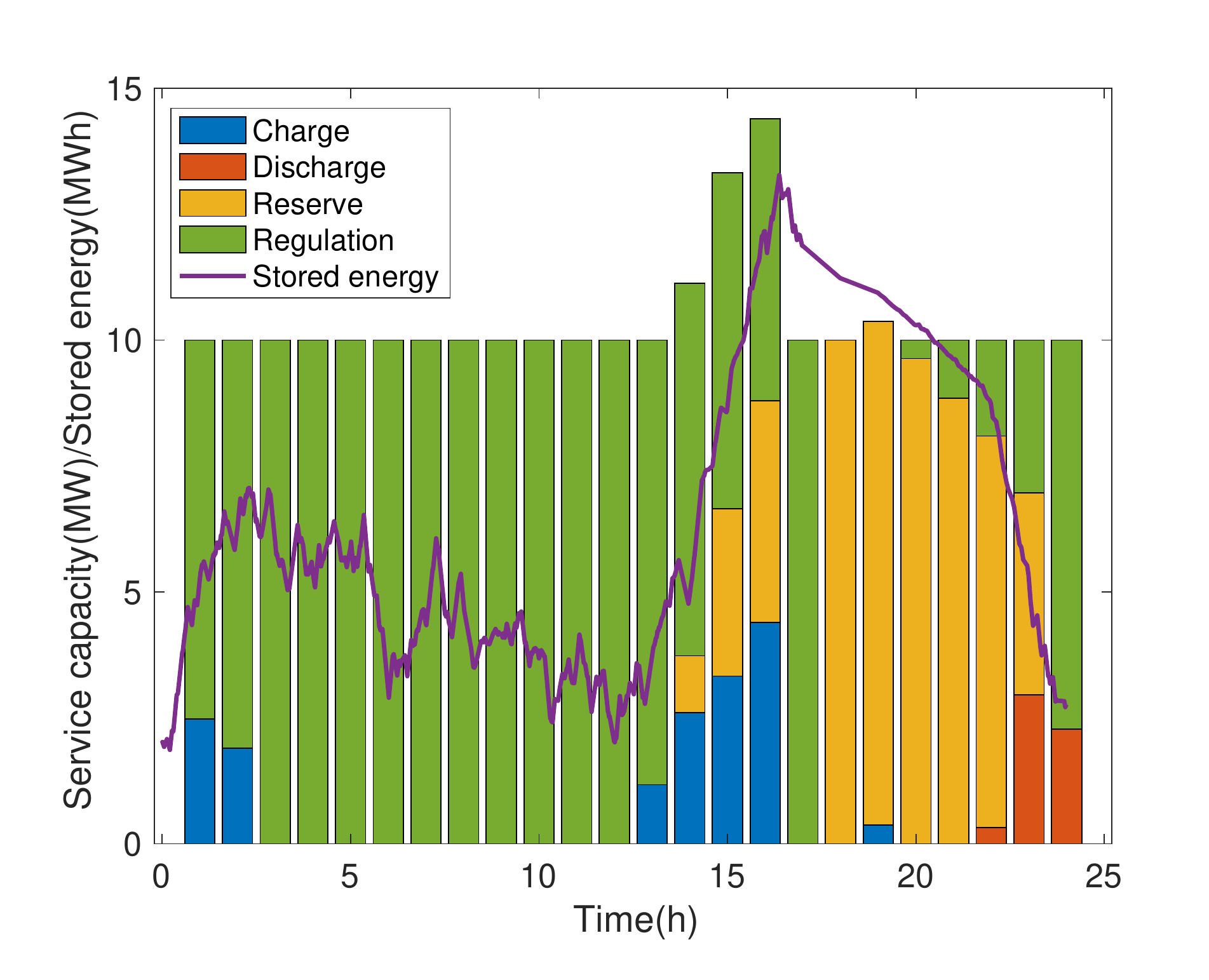}
  \caption{The single-day operation simulation result of LFP battery storage example project on days in cluster 4 in 2021}
  \label{fig:cluster4_ope}
\end{figure}

Figure \ref{fig:cluster4_ope} shows the operation simulation result of cluster 4. During hours 1-12, in which the revenue for frequency regulation is relatively high, the battery storage provides frequency regulation service following RegD, resulting in "burrs" on the energy curve. Due to the low LMP in hours 1-2, the battery storage is charged within hours 1-2 to support the energy demand afterward. During hours 13-16, in which LMP is still relatively low while the income of frequency regulation reduces, the battery storage is charged to support the energy demand for subsequent frequency regulation and spinning reserve services, while the remaining capacity is used to provide frequency regulation and spinning reserve services. During hours 18-22, in which the revenue of spinning reserve is relatively high, the battery storage mainly provides spinning reserve service. From hour 23 to hour 24, the battery storage is discharged to ensure that the energy stored in the battery returns to the initial value at the end of the day.

The single-day revenue and cost of LFP battery storage example project in cluster 4 situation is concluded as table \ref{tab:eg1_LFP_rev}. The frequency regulation revenue based on the performance-based regulation mechanism reaches 97.57\%, which contributes the most to the revenue of battery storage. Due to the small price difference between peak and off-peak of LMP curve and the charging demand to support auxiliary services, the benefit from load shifting is presented as a negative number. The single-day operation and maintenance cost is 1.75\% of the service revenue, and the capacity of energy storage decreases by 0.0154\%, corresponding to the degradation cost which accounts for 2.69\% of the service revenue of the day.

\begin{table}[H]
\renewcommand{\arraystretch}{1.3}
\centering
\caption{The single-day revenue and cost of LFP battery storage example project on days in cluster 4}
\label{tab:eg1_LFP_rev}
    \scalebox{0.9}{\begin{tabular}{p{3.5cm} p{2cm} p{2cm}}
    \toprule
    Type & Revenue/Cost & Portion\\
    \midrule
    Revenue & \$8,650 & 100\%\\
    \midrule
    Load shift & \$-443 & -5.12\%\\
    Frequency regulation & \$8,440 & 97.57\%\\
    Spinning reserve & \$653 & 7.55\%\\
    \midrule
    Maintenance and operation cost & \$-151 & -1.75\%\\
    \midrule
    Capacity loss & 0.0154\% & N.A.\\
    Degradation cost & \$-233 & -2.69\%\\
    \bottomrule
    \end{tabular}
    }
\end{table}

\subsection{Method Comparison with LCOS}\label{section5.3}
As mentioned in section \ref{section4.2}, this paper conducts the method of selecting 36 typical days in each year and the clustering method to calculate the IRR of the battery storage project. To better embody the advantages of IRR, we also apply the LCOS method in comparison.

 According to equation \ref{eq:LCOS}, in which $W_{out,t}$ is obtained by making a weighted summation to the data of 5 clusters in the clustering method which is introduced in detail in section \ref{section5.3}, the LCOS of LFP and NCM example projects are calculated as equation \ref{eq:LCOS_result}. The discount rate i is reasonably set as 8\% between 2022 and 2026, 7\% between 2027 and 2031, and 6\% between 2032 and 2036 in reference to \cite{lai2017levelized}. 
 
\begin{subequations}
    \label{eq:LCOS_result}
\begin{align}
\textit{LCOS}_\textit{{LFP}} & = 55.42\$/\textit{MWh}\\
\textit{LCOS}_\textit{{NCM}} & = 62.66\$/\textit{MWh}
\end{align}
\end{subequations}
Through the LCOS method, we conclude that LFP is more economically friendly than NCM. However, since the LCOS method only presents the full life cycle cost for battery providing per MWh of power, it is only possible to compare the cost with the price difference between peak and off-peak of the LMP curve to roughly estimate the profit from load shifting. It is difficult for the LCOS method to analyze the economic viability when the battery storage participates in frequency regulation and spinning reserve, even though the revenue from auxiliary service revenue accounts for a large portion of the battery storage revenue. Besides, since LCOS is mainly used to compare the quantitative difference in life cycle economic viability between different types of battery storage, it has relatively low accuracy.

It is calculated that the LFP battery storage example project can run for about 14 years before the capacity decreases to 80\% of the initial capacity. Therefore, for the method of selecting 36 typical days in each year, it is assumed that the project will be invested in 2021, put into operation in 2022, and decommissioned in 2035. Since it is calculated that the capacity loss of the clustering method is less than that of the method of 36 typical days possibly due to the overestimate of capacity loss when simply magnifying decuple as the result of a year, it is reasonably assumed that the LFP example project will be decommissioned in 2036 when the clustering method is applied. 

The cash flow of the project in each year of the whole life cycle is calculated from the costs at the beginning of the investment period, the service income as well as operation and maintenance costs of each year during the operation period, and the recycling income at the end of the investment period. When adopting the clustering method, it is calculated that the cash flow of 2021 is a negative number due to the initial investment of the project, including a battery purchasing cost of \$4,149.9k, a equipment purchasing cost of \$1,946.2k, and a power station design and construction cost of \$1,283.7k. In 2022, the battery storage earns a revenue of \$3,081.4k by providing auxiliary services and participating in the energy market with a maintenance and operation cost of \$57.2k. Since its revenue overweighs the maintenance and operation cost, the cash flow of 2022 is presented as a positive number. The cash flow of 2023-2035 is calculated similarly. In 2036 when the battery is decommissioned, the battery storage project earns a extra recycling income aside from the service revenue and the maintenance and operation cost, reflected as a higher cash flow this year. Therefore, the IRR of this LFP example project is calculated to be 40.78\% when the method of clustering is adopted. Through similar calculation, it is obtained that the IRR result using the method of 36 typical days is 35.90\%.  

Similarly, it is obtained that NCM battery storage example project can run for about 12 years before the cumulative battery capacity decreases to 80\% of the initial capacity. Therefore, for the method of selecting 36 typical days in each year, it is assumed that the project is invested in 2021, put into operation in 2022 and decommissioned in 2033, while the project is assumed to be decommissioned in 2034 when using clustering method. In a similar way of calculation to that of LFP example project, it is calculated through the method of 36 typical days that IRR of the NCM example project is 21.49\% , while the clustering method presents the IRR result of 25.07\%.

\begin{figure}[htbp]
  \centering
  \includegraphics[width=0.85\linewidth]{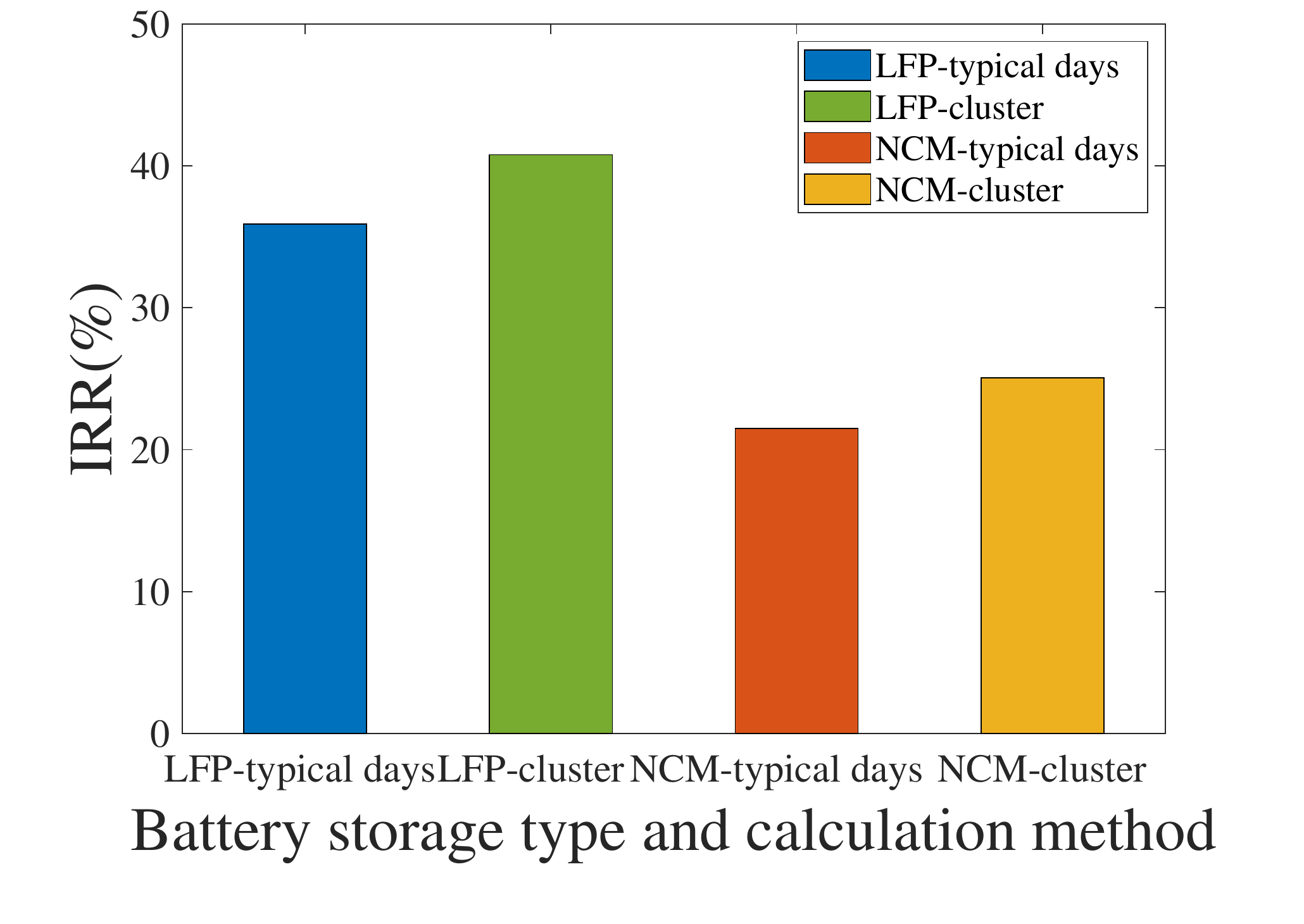}
  \caption{The IRR results of LFP and NCM battery storage project calculated through two different methods}
  \label{fig:IRR_figure}
\end{figure}

The IRR results of LFP and NCM battery storage project applying the method of selecting 36 typical days and the clustering method are summarized in figure \ref{fig:IRR_figure}.

Therefore, we draw to the conclusion that the IRR calculated through the clustering method is higher than that calculated using the data of typical days both for the LFP and NCM projects. The possible reason for this is that the days with higher electricity prices, which are clustered as a type in the clustering method, are not chosen as the typical days. The randomness in selecting typical days results in its lower accuracy than the clustering method.

From the case study above, it is concluded that both methods give positive predictions for the battery storage project of the next 15 years. The clustering method shows that the IRR of LFP and NCM battery storage example projects reach 40.78\% and 25.07\% respectively, which is relatively high in comparison to Chinese ten-year treasury bill yields since 2015 which fluctuates between 2.5\% and 4.0\%. Besides, from the perspective of economy, LFP will be a better choice for large-scale power generation side and grid-side battery storage due to its lower price and longer life cycle.

The significant economic viability of investing in a battery storage project at present is mainly due to the fact that the current price of battery energy storage has dropped to a more reasonable range, which has greatly reduced the initial investment cost. Apart from that, the gradually mature electricity market mechanism also results in more reasonable revenue for battery storage's participation in the electricity market. In the future, the increasing battery recycling income will further improve the life cycle economic viability of battery storage.
\subsection{Discussion on Capacity Degradation}
As mentioned in section \ref{section3.1}, it is set in this paper that the battery should be replaced when its capacity decays to 80\% of the initial capacity. Although 80\% is a standard commonly applied for lithium-ion batteries used in electric vehicles instead of grid-side battery storage, considering that reference \cite{reniers2021unlocking} also used 80\% for research on the grid-side applications, 80\% might be a relatively conservative value used for academic research on grid-side battery storage application. Instead, there is no fixed standard for capacity degradation limitation of grid-side battery storage application at present, and it is suggested that replacing the lithium-ion battery for grid application when its capacity decays to 60\%-70\% of the initial capacity would be a reasonable choice. Therefore, this paper calculates the IRR of LFP and NCM battery storage example project with the clustering method adopted when the capacity degradation limitation is set as 60\% of the initial capacity. Considering that battery capacity will decrease at an accelerated rate after decaying to 80\%, which is not included in the model, it is assumed that the capacity loss after the battery capacity decays to 80\% is 3 times the calculated value. Under this assumption, it is found that the life of LFP and NCM battery storage example projects are 20 years and 18 years respectively, and the IRR of LFP and NCM battery storage example projects reach 40.93\% and 25.73\% respectively. Comparing with the results when the capacity degradation limitation is set as 80\%, which are 40.78\% and 25.07\% for LFP and NCM battery storage project respectively, it is concluded that the difference is relatively small.

In equation \eqref{eq:E_max}, the value of upper energy limit $E_{max}$ is set at 90\% of the initial capacity. Although $E_{max}$ should reduce due to capacity degradation, considering that the revenue from frequency regulation is the main source of revenue for battery storage in grid-side application, the revenue of battery storage depends more on its ramping capability that makes it a regulation resource of higher quality, rather than its maximum capacity. Therefore, the error caused by capacity degradation of setting $E_{max}$ as a constant may not have a significant impact on the total revenue. To support this assertion, IRR of LFP battery storage example projecs is recalculated to be 40.71\% with the clustering method adopted when $E_{max}$ is set at 60\% of the initial capacity, which is close to the IRR result of 40.78\% when $E_{max}$ is set at 90\% of the initial capacity, and recalculation for IRR of NCM battery storage example project also come to similar conclusion. Therefore, it is concluded that setting $E_{max}$ as a constant value would be reasonable without large error.

\subsection{Sensitivity Analysis}
Considering the uncertainty of electricity market price $\textit{Price}^{\textit{e,reg,res}}$, the battery 100\% equivalent cycles $N_{100\%}$, and the decrease ratio of battery purchasing cost $k_{dec}$ in the coming 15 years, this paper conducts sensitivity analysis for LFP and NCM battery storage example projects to further estimate the life cycle economic viability.
\begin{table}[H]
\renewcommand{\arraystretch}{1.3}
\centering
\caption{Sensitivity analysis of LFP battery storage example project}
\label{tab:sen_LFP}
    \scalebox{0.7}{\begin{tabular}{c c c c c c c}
    \toprule
     & $Income$/\$K & $\textit{Cost}^{op}$ =/\$K & $ \textit{Cost}^{loss}$/\$K \\
    \midrule
    Standard case & 3081.4 & 57.2 & 97.0 \\
    Price increases 10\% & 3390.6 & 57.3 & 97.9 \\
    Price increases 5\% & 3236.4 & 57.3 & 97.8 \\
    Price decreases 5\% & 2925.4 & 56.9 & 95.4 \\
    Price decreases 10\% & 2771.2 & 56.8 & 95.2 \\
    {\footnotesize100\%-DOD cycle number decreases 50\%}& 3070.7 & 56.1 & 181.3 \\
    {\footnotesize100\%-DOD cycle number decreases 25\%}& 3077.7 & 56.7 & 125.6 \\
    {\footnotesize100\%-DOD cycle number increases 10\%} & 3082.3 & 57.3 & 89.0 \\
    {\footnotesize Battery cost decrease ratio is $6/11$}& 3077.3 & 56.6 & 140.8 \\
    {\footnotesize Battery cost decrease ratio is $2/11$}& 3089.8 & 58.5 & 53.4 \\
    \midrule
    & $Cap^{loss}$ /\% & lifetime/year & IRR/\% \\
    \midrule
    Standard case & 6.43\% & 15 & 25.07\% \\
    Price increases 10\% & 6.49\% & 15 & 27.93\% \\
    Price increases 5\% & 6.48\% & 15 & 26.51\% \\
    Price decreases 5\% & 6.32\% & 16 & 23.62\% \\
    Price decreases 10\% & 6.31\% & 16 & 22.16\% \\
    {\footnotesize100\%-DOD cycle number decreases 50\%}& 12.01\% & 8 & 20.35\% \\
    {\footnotesize100\%-DOD cycle number decreases 25\%}& 8.32\% & 12 & 23.72\% \\
    {\footnotesize100\%-DOD cycle number increases 10\%}& 5.90\% & 17 & 25.32\% \\
    {\footnotesize Battery cost decrease ratio is $6/11$}& 6.22\% & 16 & 25.19\% \\
    {\footnotesize Battery cost decrease ratio is $2/11$}& 8.187.08\% & 14 & 24.87\% \\
    \bottomrule
    \end{tabular}
    }
\end{table}

\begin{table}[H]
\renewcommand{\arraystretch}{1.3}
\centering
\caption{Sensitivity analysis of NCM battery storage example project}
\label{tab:sen_NMC}
    \scalebox{0.7}{\begin{tabular}{c c c c c c c}
    \toprule
     & $Income$/\$K & $\textit{Cost}^{op}$ =/\$K & $ \textit{Cost}^{loss}$/\$K \\
    \midrule
    Standard case & 3066.1 & 55.8 & 232.1 \\
    Price increases 10\% & 3374.3 & 55.9 & 233.5 \\
    Price increases 5\% & 3220.6 & 55.8 & 233.2 \\
    Price decreases 5\% & 2911.4 & 55.6 & 230.8 \\
    Price decreases 10\% & 2757.0 & 55.6 & 229.6 \\
    {\footnotesize100\%-DOD cycle number decreases 50\%}& 3032.5 & 53.7 & 420.5 \\
    {\footnotesize100\%-DOD cycle number decreases 25\%}& 3055.2 & 55.2 & 298.1 \\
   {\footnotesize100\%-DOD cycle number increases 10\%}& 3067.6 & 55.9 & 212.3 \\
    {\footnotesize Battery cost decrease ratio is $6/11$}& 3052.4 & 55.0 & 332.6 \\
    {\footnotesize Battery cost decrease ratio is $2/11$}& 3077.8 & 56.7 & 123.3 \\
    \midrule
    & $Cap^{loss}$ /\% & lifetime/year & IRR/\% \\
    \midrule
    Standard case & 7.69\% & 13 & 25.07\% \\
    Price increases 10\% & 7.74\% & 13 & 27.93\% \\
    Price increases 5\% & 7.73\% & 13 & 26.51\% \\
    Price decreases 5\% & 7.65\% & 13 & 23.62\% \\
    Price decreases 10\% & 7.61\% & 13 & 22.16\% \\
    {\footnotesize100\%-DOD cycle number decreases 50\%}& 13.93\% & 7 & 20.35\% \\
    {\footnotesize100\%-DOD cycle number decreases 25\%}& 9.88\% & 10 & 23.72\% \\
    {\footnotesize100\%-DOD cycle number increases 10\%}& 7.03\% & 14 & 25.32\% \\
    {\footnotesize Battery cost decrease ratio is $6/11$}& 7.35\% & 14 & 25.19\% \\
    {\footnotesize Battery cost decrease ratio is $2/11$}& 8.18\% & 12 & 24.87\% \\
    \bottomrule
    \end{tabular}}
\end{table}

The following conclusions are obtained from the sensitive analysis:

(a) The fluctuation of $\textit{Price}^{\textit{e,reg,res}}$ has remarkable influence on annual service revenue $\textit{Income}$. $Cap^{loss}$ also increases with the rise of $\textit{Price}^{\textit{e,reg,res}}$ due to the more aggressive bidding strategy incentivized by the higher service revenue. Yet the influence on $Cap^{loss}$ is insignificant and the lifetime of battery almost remain the same with the fluctuation of $\textit{Price}^{\textit{e,reg,res}}$. As a result, IRR has a significant positive correlation with $\textit{Price}^{\textit{e,reg,res}}$.

(b)The reduction of $N_{100\%}$ has significantly influenced $Cap^{loss}$, thus increasing the degradation cost $\textit{Cost}^{loss}$ and shortening the lifetime of the battery, which also greatly influences IRR.

(c)$k_{dec}$ decides the battery replacement cost, which remarkably influences $\textit{Cost}^{loss}$, resulting in different bidding strategy. $Cap^{loss}$ and battery lifetime change slightly correspondingly. Therefore the decline of $k_{dec}$ also slightly causes the rise of IRR. 

(d)Since the fixed part account for a large portion in $\textit{Cost}^{op}$, the fluctuation of $\textit{Price}^{\textit{e,reg,res}}$, $N_{100\%}$, and $k_{dec}$ has little influence on $\textit{Cost}^{op}$.

In conclusion, $\textit{Price}^{\textit{e,reg,res}}$ and $N_{100\%}$ have significant impact on the economic viability of battery storage. The rise in annual service revenue and the battery lifetime will greatly benefit the development of battery storage in the future.
\section{Conclusion}
An accurate and comprehensive analysis of life cycle economic viability is vital to the development of battery storage. Therefore, this paper proposes a life cycle economic viability analysis model with the variable revenue and cost calculated through a battery storage operation simulation model considering degradation. 
The fixed cost and revenue which depend on the specific battery storage project are also included in the economic viability analysis model. The income and cost of the variable and fixed parts of battery storage can be aggregated to obtain the cash flow of each year in its entire life cycle, and IRR can be calculated accordingly. A method selecting 36 typical days each year and a clustering method based on the Kmeans algorithm with dynamic time warping metric are used to conduct case studies respectively, in which the clustering method is found to be more accurate. It is calculated that the IRR of LFP and NCM battery storage projects are 40.78\% and 25.07\% respectively at present, which is relatively high mainly due to the declining battery cost and the increasingly mature electricity market mechanism. Through sensitivity analysis, it is found that electricity price and DOD have a significant impact on the economic viability of battery storage, which indicates that the economic viability of battery storage will be greatly improved when annual service revenue and battery lifetime increase.











\printcredits

\bibliographystyle{unsrt}




\end{document}